\documentclass[dvips]{acta}
\usepackage{supertabular,lscape,epsfig}
\usepackage{amssymb}
\usepackage{amsmath}
\usepackage[dvipsnames]{xcolor}
\usepackage{wasysym}
\usepackage{pdflscape}
\SetPages{0}{0}
\graphicspath{ {image/} }
\SetVol{68}{2018}

\begin{document}

\begin{Titlepage}
\Title{Hybrid Pulsations and Tidal Splitting detected in the {\it Kepler} Eclipsing and Spotted Binary System KIC 6048106}
\Author{Samadi Gh.$^1$, A. and Lampens$^2$, P. and M. Jassur$^3$, D.}
{$^1$Research Institute for Applied Physics and Astronomy, University of Tabriz, Tabriz, Iran\\
e-mail:samadi.aunia@tabrizu.ac.ir\\
$^2$Koninklijke Sterrenwacht van Belgi\"e, Ringlaan 3, B-1180 Brussels, Belgium\\
$^3$Department of theoretical Physics and Astrophysics, Physics Faculty, University of Tabriz, Tabriz, Iran}

\Received{May 27, 2018}
\end{Titlepage}

\Abstract{We present a new asteroseismic analysis of KIC~6048106, a \textit{Kepler} Algol-type eclipsing binary star in a circularized orbit with $P_\mathrm{orb}$=1.559361$\pm$0.000036~d. Based on a physical model for the binary  and its corresponding set of fundamental parameters, ($T_\mathrm{eff}=7033\pm187~K, ~M_\mathrm{1}=1.55\pm0.11M_{\odot},~R_\mathrm{1}=1.58\pm0.12R_{\odot}$ and $T_\mathrm{eff}=4522\pm103~K, ~M_\mathrm{2}=0.33\pm0.07M_{\odot},~R_\mathrm{2}=1.78 \pm0.16R_{\odot}$, respectively for the primary and the secondary component), we obtained the residual light curve after removal of the full binary model, including a 290-day activity cycle for the secondary component (Samadi Gh. et al.\,2018). In this work, we used the method of Fourier analysis of the residual light curve in combination with least squares optimization for the frequency analysis. We detected seven dominant, independent gravity ($g$) modes as well as 34 low-amplitude acoustic ($p$) modes. The $g$ modes in the range $1.96-2.85$ d$^{-1}$ have a mean spacing of $\Delta\Pi_\mathrm{mean}=1517.92\pm131.54$~s. Though of much lower amplitude, additional significant frequencies were detected in the intervals $7.49-15.2$ d$^{-1}$ and $19-22.5$ d$^{-1}$ (i.e. in the $p$ mode region), with corresponding dominant modes $\nu_\mathrm{max_\mathrm{1}}=11.745\pm0.001$ d$^{-1}$ and $\nu_\mathrm{max_\mathrm{2}}=20.960\pm0.002$ d$^{-1}$. From its position in the H-R diagram, we conclude that the primary component is the source of the detected hybrid pulsations. Consequently, the pulsation constants, Q, of the high frequencies cover the range $0.028 - 0.064$~d. Furthermore, $\nu_\mathrm{43} $ (19.037$\pm$0.002 d$^{-1}$) might correspond to the fundamental radial mode (Q = 0.033$\pm$0.007 d). The other frequencies in the range $19-22.5$ d$^{-1}$ could be radial or non-radial overtone modes. Moreover, the low-amplitude $p$ modes show an equidistant splitting by $f_\mathrm{orb}$, which we interpret as tidal splitting following theoretical predictions.} 
{techniques: photometric --
stars: binaries: eclipsing --
stars: fundamental parameters --
stars: variables: delta Scuti --
stars: oscillations -- 
stars: starspots}

\section{Introduction}
The intermediate-mass A-F type stars are in a phase of transition when located near the red edge of the $\delta$ Scuti instability strip. This transition occurs when the thin convective zones of the H and He~II partial ionization zones merge to form a single, larger and more efficient convective zone below the stellar surface (Dupret et al.\,2007). The $\delta$ Scuti and $\gamma$ Doradus stars are found among the A-F type stars undergoing such changes of their structure. \\
$\delta$ Scuti stars are (pre-, normal or post-) main-sequence A-F type stars with masses and effective temperatures in the ranges of respectively 1.5 to 2.5$M_{\odot}$ and 6900 and 8900 K (on the main sequence) which pulsate in low-order radial or non-radial pressure ($p$) modes (Aerts, Christensen-Dalsgaard \& Kurtz 2010). Such modes are sensitive to the stellar surface layers, and their restoring force is pressure. Their periods range from 15~min to 6~hours.  \\
$\gamma$ Doradus stars are the less massive (1.5-1.8$~M_{\odot}$), slightly cooler (6700-7400 K) counterparts of $\delta$ Scuti stars and present both photometric and line profile variations (Catelan et al.\,2015). They exhibit high-order, low-degree multiperiodic gravity ($g$) mode pulsations. These $g$ modes occur with periods of 7 hours to 3 days.  \\
The $\gamma$ Doradus instability strip overlaps the cool part of the $\delta$ Scuti instability strip.  Concerning the excitation mechanisms, Xiong et al.\ (2016) reported that the $\kappa$ mechanism and coupling of convection and oscillations are driving the pulsations in the hot and cool $\delta$ Scuti/$\gamma$ Dor stars, respectively.\\
For stars located in the overlap region of the $\gamma$ Dor and the $\delta$ Scuti instability strips, we expect to find both low- ($g$) and high-frequency ($p$) pulsation modes. According to Balona et al.\ (2015), all the \textit{Kepler} $\delta$ Scuti stars are hybrid pulsators. The hybrid pulsators provide information from the surface ($p$ modes) as well as from the near-core regions ($g$ modes). When such a pulsator is part of an eclipsing binary, the situation becomes even more favorable for a deep analysis: by means of light curve modeling of the system's properties, one can obtain model-independent and (more) accurately determined fundamental stellar parameters (than in the case of single stars), therefore more reliable constraints for stellar models. Additionally, the information derived from a detailed pulsation study (\textit{i.e.} $ \nu_\mathrm{max},~\Delta \nu $ and individual pulsation frequencies) can be associated to fundamental stellar parameters like the mean stellar density $\overline{\rho}$ (\textit{e.g.} Su\'{a}rez 2014). These two independent approaches (a light curve analysis and its associated binary modeling to determine the component properties and the complementary frequency analysis of the residual light curve) allow to revisit the stellar evolution models, in connection with the asteroseismic modeling. \\
Nowadays, space missions like \textit{Kepler} (and \textit{K2})~(Koch et al. 2010) are pioneer in providing ultra-high-accuracy photometry to study stellar pulsations as well as other stellar mechanisms. The continuous photometric data acquired with an accuracy of the order of $\mu$mag is suitable for detecting high-order, low-degree $g$ modes and low-order $p$ modes, which have amplitudes (sometimes) lower than 0.1 mmag. Examples of recent studies on binary systems with hybrid pulsations (in one or both components), are He{\l}miniak et al.\ (2017) and Guo et al.\ (2017). The former concerns a study of hybrid pulsations in a \textit{Kepler} quintuplet system, KIC~4150611, with two pairs of eclipsing binaries. The brightest component in the system (Aa) shows hybrid pulsations. The latter work concerns the study of a post-mass transfer hybrid pulsator in the eclipsing binary KIC~9592855 and reports that both core and envelope rotation rates are similar to the orbital frequency. Lampens et al.\ (2018) investigated the binary fraction among 50 \textit{Kepler} A-F-type candidate hybrid variable stars. Based on multi-epoch high-resolution spectra collected between 2013 and 2016, the authors find that 27\% of their sample stars are part of a binary or triple system.  \\
Here, we present a pulsation study of KIC~6048106, an Algol-type \textit{Kepler} eclipsing binary system with K$_{\mathrm p}$ of 14.1 mag. From the light curve modeling, based on a physical binary model, Samadi Gh. et al.\ (2018) derived the orbital properties of the system and stellar fundamental parameters of both components. The Eclipse Time Variations (ETVs) for both components were also analyzed. After having 
completed the analysis of the residuals, we became aware of the study reported by Lee (2016). It is essential to confront both analyses, in particular with respect to the definition of the binary model in the first place. We will deal with this in Section 2, and argue that our model is more consistent. Consequently, the frequency-analysis of the (improved) residual signal will reveal further information in our case (Section~4.3).  \\
We will review the characteristics of the \textit{Kepler} observations for our target in Section~2. The methodology used for deriving the stellar parameters and its results is also presented. The methodology used for the frequency analysis along with the theoretical proof of the detected frequency splitting are described in Section~3. The details of the Fourier analysis and the detection of the $g$ and $p$ modes are provided in sections~4.1,~4.2, and 4.3, respectively. We discuss the pertinence of the high-frequency modes in Section~4.4. Finally, we mention the outcome of a Fourier analysis of the final residuals in Section~4.5. The overall results and conclusions are presented in Section~5.
\section{\textit{Kepler} Photometry, Methodology and the Binary Modeling}
KIC~6048106 is a short-period eclipsing binary system with an orbital period, $P_\mathrm{orb}$, of 1.559361 $\pm$ 0.000036 d\footnote{\textit{Kepler} Eclipsing Binary Catalogue (http://keplerebs.villanova.edu/)}. The available data consist of two Long Cadence (LC) light curves from quarters 14 and 15 (with a sampling of 29.42 min). These quarters cover a 197-days observation period from 2012-06-28 to 2013-01-11. There are some significant events that resulted in gaps in the light curves of both Q14 and Q15. The gaps occur during the orbital cycles 828-830, 880, 907-909. The light curves were first smoothened using a 2nd-order polynomial to fit the mean out-of-eclipse flux level, for each quarter separately. Next, the data from both quarters were converted to the magnitude scale and concatenated. We used this \textit{detrended Kepler} light curve for the current study. \\
Before calculating the best-fitting binary model, Samadi Gh. et al.\ (2018) estimated the contribution of the pulsations in the light curve. They produced a multi-parameter fit that was obtained by using all the frequencies with a Signal-to-Noise Ratio (SNR)$>4$ d$^{-1}$ (Breger et al. 1993) (using an iterative prewhitening process) extracted from the detrended \textit{Kepler} light curve (excluding the known orbital frequency, $f_{orb}$ = 0.64143 $\pm$ 0.00015 d$^{-1}$,  and its harmonics). Fig. 2 (top left panel) illustrates the Fourier spectrum of the detrended light curve (top panel in Fig. 1). It is obvious that the orbital frequency and its harmonics are the most dominant frequencies, with amplitudes of the order of tens of mmag. However, there are also some additional significant frequencies, other than the orbital harmonics, both in the low- and the high-frequency regions (Table~A1). By subtracting the resulting multi-parameter fit from the detrended light curve, they obtained a residual light curve from which the pulsational contribution was approximately removed. They then searched for the best-fitting binary model (with minimum $\chi^{\mathrm 2}$) in order to accurately determine the stellar and orbital parameters using \verb'PHOEBE' (Pr\v{s}a et al. 2005). Because of significant and easily detectable light modulation at both quadrature phases, they also included a cold spot (R = 11.0R$_{\odot}$ and $\frac{T_{spot}}{T_{surf}}$ = 0.8) on the surface of the secondary star. Samadi Gh. et al.\ (2018) thus derived a consistent binary model including also a variable spot model, going from a low (L), through a medium (M) to a high (H) state of stellar activity (Tables 5 and 6 in Samadi Gh. et al.\ 2018). Since the secondary component is a cool $\sim$K3-K4 subgiant, they considered as very probable that (most of) the spots are located on its surface rather than on that of the primary component.\\
Moreover, from an ETV modeling for the secondary star, they detected a trend with a period of $ P_\mathrm{mod} = 278\pm7$ d that could be related to magnetic activity (according to the Applegate (1992) mechanism).\\ 
Samadi Gh. et al. (2018) looked at the behavior in time of all the residuals and defined three segments of the light curve with a clearly distinct pattern in the residuals (their Fig. 8) which were attributed to different phases of activity (spot models). They also found a long-term cyclic variation causing brightness fluctuations in the residuals of the best-fitting binary model. Moreover, they independently confirmed the modulation of the maximum light which indicates a period $ P_\mathrm{mod}$ = 290$\pm$7 d. This observation is consistent with the presence of a similar modulation in the {\it Kepler} ETV of the secondary component (with $ P_\mathrm{mod} = 278\pm7$ d). In summary, KIC 6048106 is a semi-detached eclipsing binary system with a circular orbit, slight asynchronous rotation and orbital inclination $i = 73.30^{\circ}\pm2^{\circ}$ (Samadi Gh. et al.\,2018). \\
On the other hand, Lee (2016) used a step-by-step fitting which included changing the properties of a single spot (his Tables 1 and 2) based on 113 light curve segments (\textit{i.e.} one segment for each orbital cycle) to consider the variation of the light level at the quadrature phases. Though the spot models are different, the stellar and the orbital parameters derived from both studies are in all aspects equivalent. However in Samadi Gh. et al. (2018)'s case, the mean level of the residuals is 2.8 mmag (their Fig. 2), whereas in Lee's case, the mean level of the residuals (considered over a period of 20 days) is at least twice as high (Fig. 4 in Lee 2016). Thus, the residual light curve, under analysis here, is about two times smoother. 
\section{Frequency analysis}
We used the Lomb-Scargle method (Lomb 1976, Scargle 1982) for the frequency analysis, up to Nyquist frequency, $f_\mathrm{Ny}\simeq$ 24.65 d$^{-1}$. We calculated the noise spectrum after prewhitening of the peak with the highest SNR, identified the next peak of highest SNR, and applied a convolution in a 1d$^{-1}$ window. The errors on the frequencies (and their phases and amplitudes) were calculated following Montgomery \& \'{O}~Donoghue (1999). We considered the correlation-correcting factor ($\sqrt{\rho} \sim $2-3) to compensate for underestimation of the errors (Schwarzenberg-Czerny 2003). For the iterative prewhitening, we used Van{\'i}{\v{c}}ek's (1971) method where the original signal is used to fit all the parameters simultaneously again.\\
In order to identify the orbital frequency and its harmonics, we considered the Rayleigh limit (here: 0.00508 d$^{-1}$) as a useful approximation of the frequency resolution ($T\geqslant 1/|\nu_{i}-\nu_{j}|,~i\neq j$; $T$ is the time span of the observations). According to Loumos \& Deeming (1978), the derived values of two neighboring (close) frequencies are inaccurate if their difference satisfies $ |\nu_{i}-\nu_{j}| \leq 2.5/T$. In such a case, we kept the one with the larger amplitude only.\\ 
A careful study of the Fourier spectrum was done in different regions where frequencies of high enough amplitude (larger than the local noise envelope) and SNR $>4$ d$^{-1}$ (Breger et al. 1993) were found. \\
According to the theory of stellar oscillations, regular spacings and simple relationships may occur between the frequencies of excited modes (for detailed explanations concerning the theory of non-radial stellar pulsations, we refer to Aerts, Christensen-Dalsgaard \& Kurtz (2010)). In the case of a genuine hybrid pulsator, since different types of  modes occur, such regularities may exist between the periods and frequencies in the $g$ and $p$ mode regions, respectively. For a useful description of the possible splittings and coupling effects, we refer to Bogn{\'a}r et al. (2015). We thus searched for regular spacings in the periods of the low-frequency peaks and the frequencies of the high-frequency peaks. This may help to infer the internal rotational profile of the pulsating component. \\
\subsection{Frequency splitting and the effects of stellar rotation}
To the first-order approximation, and given slow rotation, the frequency spacings between rotationally split components is described by:
\begin{equation}
\label{eq:rotation}
\Delta \nu_{n,\ell,m} = m (1-C_{n,\ell}) \Omega,
\end{equation}
\noindent where $C_{n,\ell}$ is the Ledoux constant whose value depends on the pulsation mode and the model, $\Omega$ is the (uniform) angular velocity (corresponding to a rotation frequency $\Omega/2\pi$) (Ledoux 1951), and $n$ is the radial order, $\ell$ is the degree, and $m$ is the azimuthal number of the spherical harmonic describing the mode (Tassoul 1980, Aerts, Christensen-Dalsgaard \& Kurtz 2010). \mbox{$C_{n,\ell} \approx 1/\ell(\ell+1)$} for high-order $g$ modes, $C_{n,\ell} \approx 0$ for high-order $p$ modes. We note that this approximation is not valid when the rotation rate becomes significant (Bouabid 2013). In the following discussion, we used the simple form of Eq. (1).
\subsection{Frequency splitting and the effects of tides}
In addition to the symmetry axis imposed by the rotation, another axis of symmetry appears when the star is a component of a close binary system undergoing tidal interaction. The companion star exerts a tidal force on the primary component whose strength can be represented as 
\begin{equation}
\label{eq:tid1}
\epsilon_{T} = (R_1/a)^3 . (M_2/M_1)
\end{equation}
\noindent where $a$ is the semi-major axis, $R_1$ the stellar radius of the primary, and $M_{1,2}$ are the component masses. The tide-generating potential can be expanded as a Fourier series, where the first term represents the  static tide and the following terms correspond to a partial dynamic tide whose frequency involves the rotational frequency and the harmonics of the orbital frequency (Smeyers 2005). The tidal potential thus contains many forcing frequencies. In close binaries, if a component rotates faster than the orbital speed ($\Omega$ > $\omega$), tidal bulges form with an angle $\simeq (\Omega - \omega$) with respect to the lines joining the star centers due to the dissipation of the kinetic energy. A restoring force tends to align the bulges along this line and the star experiences a tidal torque which pulls it into synchronization and circularization (Zahn 2005, Fig. 1). As a consequence, the stellar surfaces are no longer spherical but ellipsoidally shaped (\textit{e.g.} the $\delta$ Scuti and binary star $\theta$~Tuc, De Mey et al.\,1998, the $\delta$ Scuti and triple system DG~Leo, Lampens et al.\,2005). In the case of equilibrium tides, the spins are aligned, the rotation of both components is synchronous with the orbital speed, and the orbit is circular.\\
If one of the components is pulsating, the tidal force may cause a variety of additional effects: under the tidal action, the stellar surfaces are no longer spherical but ellipsoidal, and when synchronization is reached, the equilibrium tide generates a constant deformation. In this case, each (degenerate) eigenfrequency of a spherically symmetric star is split into ($\ell$+1) eigenfrequencies. The theoretical assumption here is that the symmetry axis of the pulsations coincides with the axis joining the two star centers. To first order in $\epsilon_{T}$, the eigenfrequencies of the radial modes ($\ell$ = 0) remain unaffected. The differences between the perturbed eigenfrequencies and the unperturbed eigenfrequency belonging to $\ell$ = 1, 2 and 3 can be computed in an arbitrarily chosen scale. The ratios of the relative differences are the same for all modes $n$ of a given degree $\ell$ (Fig. 4 in Smeyers 2005). For an observer in a non-rotating frame of reference, each (degenerate) eigenfrequency of a spherically symmetric star is then split into (2$\ell$+1) eigenfrequencies whose differences for a given $\ell$ show a regular spacing equal to (a multiple of) the orbital frequency (see Table 1 in Smeyers 2005). An example of such splitting was found for the $\delta$ Scuti and ellipsoidal star 14~Aur~A, extensively observed and studied by Fitch \& Wisniewski (1979). They detected a triplet of frequencies ($\ell$ = 1) which was explained as due to the distorted shape of the primary star, as well as doublets from tidal splitting. For a recent discussion on this topic, see Balona (2018).\\ 
In addition, in the case of an eccentric orbit, resonances may occur between a partial dynamic tide and an oscillation mode since one of the forcing frequencies happens to be close to an eigenfrequency, \textit{i.e.} the oscillation is resonantly excited. The phenomenon of tidal excitation is relevant for non-radial $g$ modes (Cowling 1941, Willems \& Aerts 2002, Willems 2003). Examples are HD~209295 (Handler et al.\,2002) as well as the pulsating heartbeat stars from the {\it Kepler} mission, \textit{e.g.} KIC~4544587 (Hambleton et al.\, 2013) and KIC~4142768 (Balona 2018).\\
In the case of an eccentric orbit and radial modes, we may detect amplitude (and phase) modulation. A good example is the $\beta$~Cephei star in a multiple system $\sigma$~Sco showing a clear modulation with $\frac{1}{4}P_\mathrm{orb}$ (Goossens et al.\,1984).  
\subsection{Coupling effects}
Coupling between $g$ and $p$ modes has already been reported in  some hybrid pulsators. Frequencies may be detected in the $p$ mode regime according to the relation $\nu_{i,p} = \nu_{max} \pm \nu_{i,g}$, where $\nu_{i,g}$ is a $g$ mode frequency and $\nu_{max}$ is the dominant $p$ mode: the fundamental or higher overtone radial modes (Bogn\'{a}r et al.\,2015). The detection of such coupling supports the idea that the frequencies found in the low- and high-frequency region originates from the same pulsator (Bogn\'{a}r et al.\,2015). Kurtz et al.\, (2014) showed how the regular spacings and their deviations can be used to probe the interior of KIC 11145123. Another example is KIC 8054146 (Breger et al.\,2012), where peaks were detected in the high-frequency region at $\nu_0 + n\nu_{max}$, with $\nu_0$ being a constant and $\nu_{max}$ being the dominant frequency in the low-frequency regime. \\
\section{Results and Discussion}
\subsection{Fourier analysis of the residual light curve}
\begin{figure}[htb]
\begin{center}
\includegraphics[width=0.8\columnwidth]{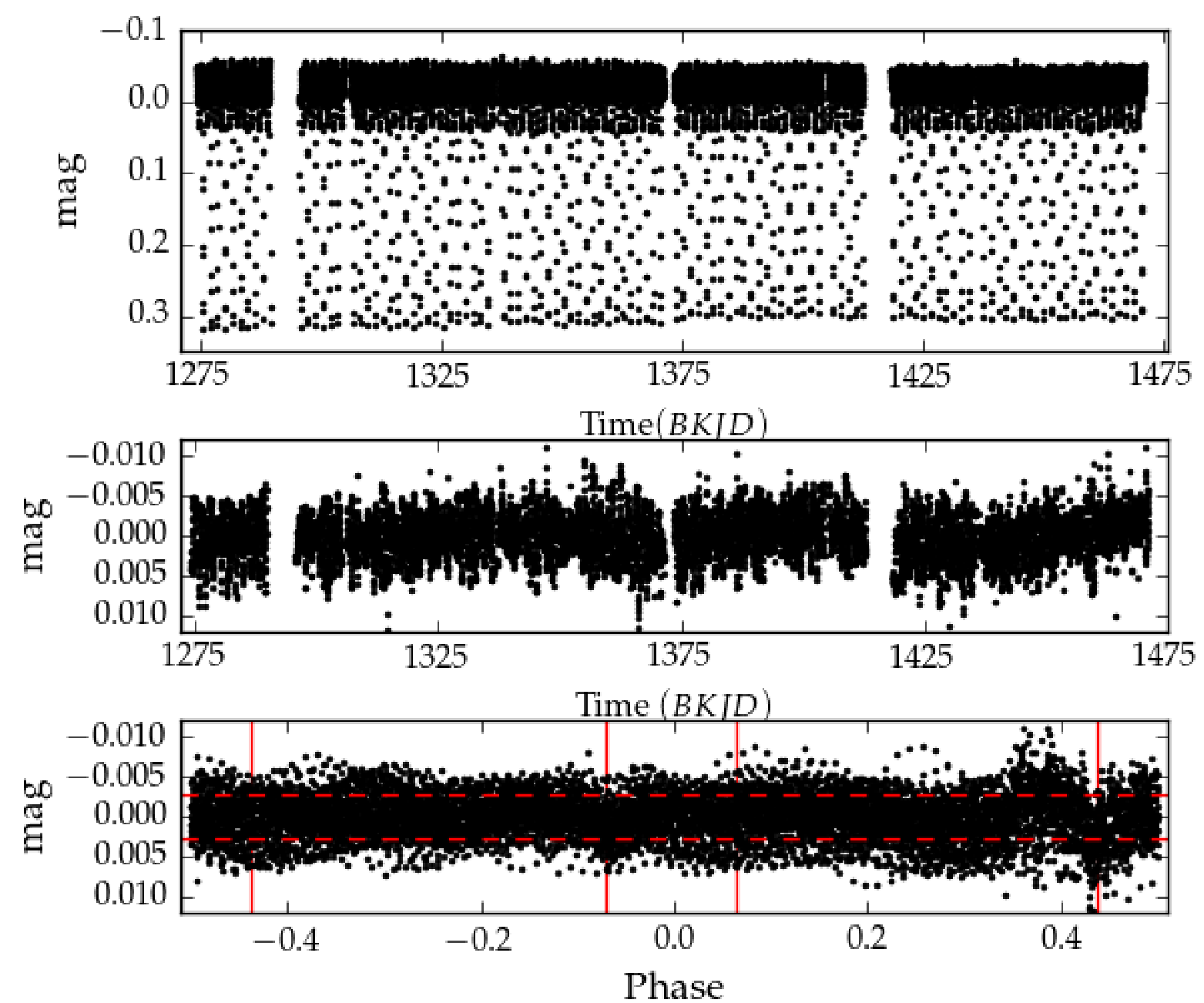}
\end{center}
\FigCap{Top: The detrended light curve of KIC 6048106. Barycentric {\it Kepler} Julian Date (BKJD) is Barycentric Julian Date (BJD) with a zero point of 2454833.0 (BKJD = BJD - 2454833.0). Middle: The residual light curve used for frequency analysis plotted against time. Bottom: The same as middle plotted against phase. The mean standard deviation of the residuals is illustrated by two horizontal red dashed boundaries ($\sigma = 0.0028$ mag, Samadi Gh. et al. 2018). The vertical solid lines indicate the phases of the primary ($ \phi $= -0.071 to 0.064) and secondary eclipses ($\phi $> 0.437 and $\phi $<-0.437) from the out-of-eclipse phases.}
\end{figure}
\begin{figure}[htb]
\includegraphics[width=\columnwidth]{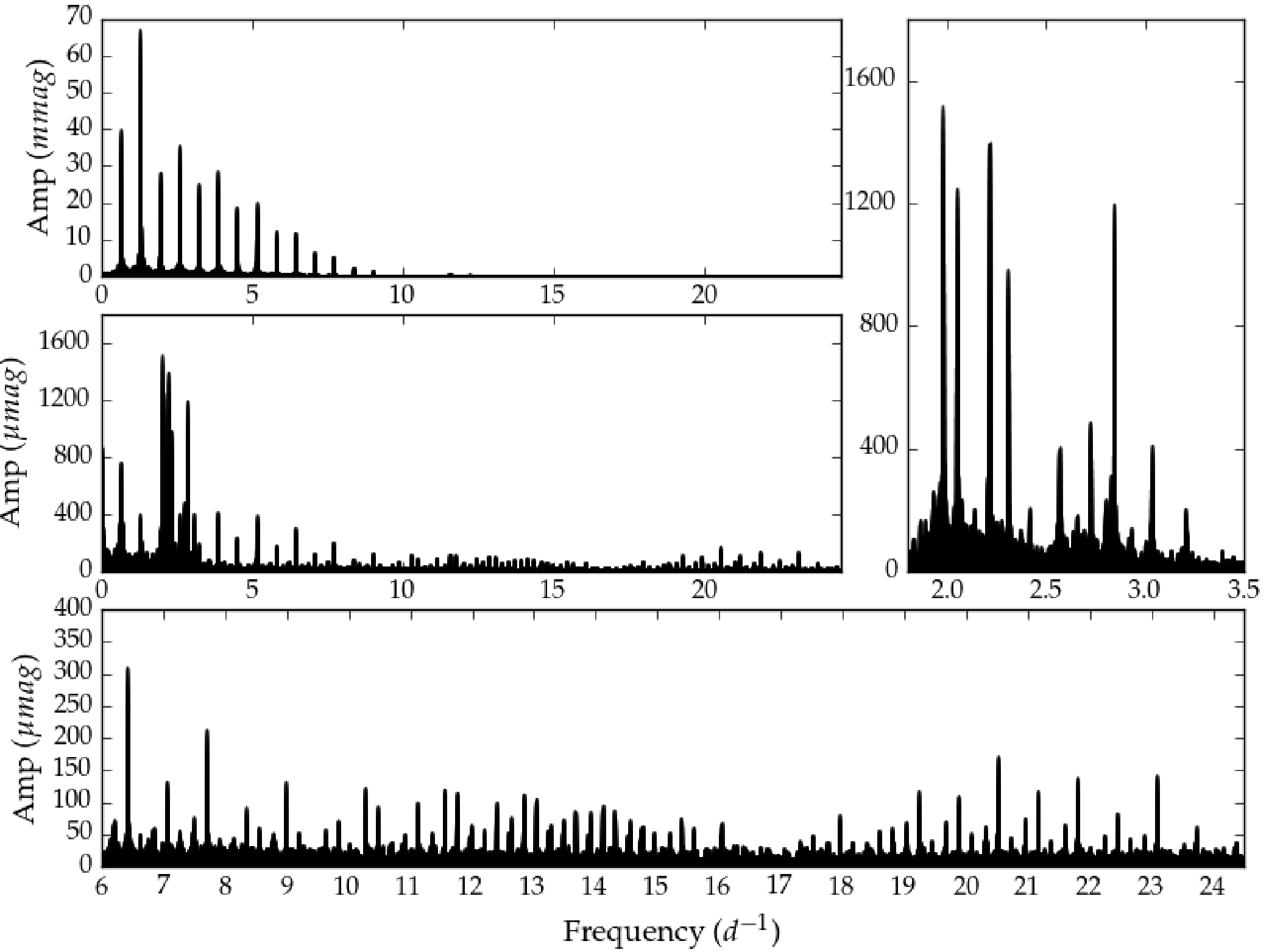}
\FigCap{The Fourier spectrum of the detrended light curve of KIC 6048106 in top panel compared with the Fourier spectrum of the residual light curve in the middle panel (the residual light curve is the residuals of the best-fitting binary model subtracted from the original light curve). Note how the amplitude of the significant frequencies decrease from mmag (in top) to $\mu$mag (in middle). The bottom and right panels are a zoom to the $p$ and $g$ mode regions of the Fourier spectrum (presented in the middle panel), respectively. The amplitudes in the right panel are in $\mu$mag.}
\end{figure}
The residuals obtained from the different states of stellar activity (high (H), medium (M) and low (L) states, cf. Section~2) were concatenated to form the residual light curve. The middle left panel in Fig. 2 shows the Fourier spectrum of the residual light curve. All the significant frequencies extracted from this Fourier spectrum are listed in Table~A2 (for the low-frequency region), Table~A3 (for the region 3.03 - 6 d$^{-1}$) and Tables~A4 and A5 (for the high-frequency region), along with their amplitudes, phases, SNR and associated errors. Most low-frequency modes are found in the frequency interval 0.6 - 3.03 d$^{-1}$. The most dominant frequency among them is $f = 1.9764\pm0.0002$ d$^{-1}$ ($A = 1600.81\pm99.07~\mu$mag, SNR = $14.38$). The only significant frequency in the region 0 - 0.6 d$^{-1}$ is $f = 0.01478\pm0.00022$ d$^{-1}$, ($A=804.49\pm62.99~\mu$mag. It is a linear combination of the form $\frac{1}{3}f_\mathrm{1}-f_\mathrm{orb}$. The frequencies in the $g$ mode region are the most significant ones (Fig. 2, right panel). Their maximum amplitude is approximately ten times larger than the maximum amplitude in the $p$ mode region (Fig. 2, bottom panel).\\
Before considering the high-frequency region of the Fourier spectrum (Fig. 2, bottom panel), we checked the frequencies in the region 3.03-6 d$^{-1}$. We know that pulsation frequencies can shift to larger frequencies under the influence of rotation (Bouabid et al.\,2013). Table~A3 shows that all significant frequencies in the frequency interval 3.03-6 d$^{-1}$ are higher harmonics of the orbital frequency. In contrast to Lee (2016), we also report the presence of two clusters of low-amplitude but significant frequencies in the regions (7-16) d$^{-1}$ and (19-24) d$^{-1}$. 
The most dominant frequency in the $p$-mode region is $11.7454\pm0.0012$ d$^{-1}$ ($A = 112.83\pm48.78~\mu$mag, SNR = $5.0$). This frequency is not an harmonic of the orbital frequency. Furthermore, the previously undetected 29$^\mathrm{th}$ orbital harmonic now appears in the Fourier spectrum.
\subsection{The low-frequency region}
\begin{figure}[htb]
\includegraphics[width=\columnwidth]{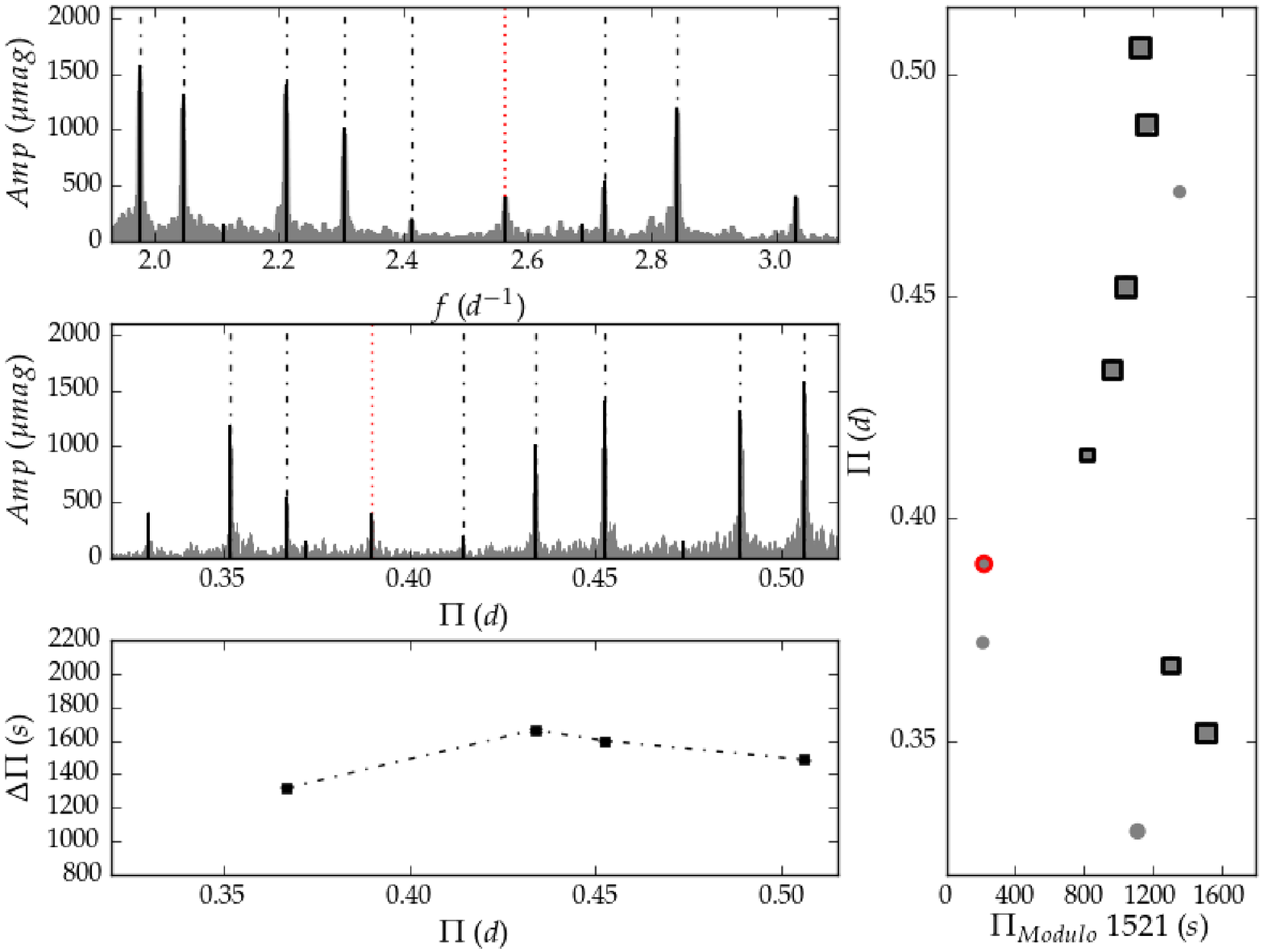}
	\FigCap{Detected $ g $ modes and their related frequencies and periods (dot-dashed lines, ".-"). Top panel shows the Fourier spectrum in the frequency regime while the middle panel shows the same in the period regime. The dotted ("...") red line designates the 3$ ^{rd} $ harmonic of the orbital frequency ($f_\mathrm{9}$ = $4f_\mathrm{orb}$). Bottom panel: detected semi-regular period spacing of $ g $ modes (Table 1). Right panel: period-\'{e}chelle diagram of the $ g $ modes ($\mu$mag). Circles in grey: all significant frequencies. Black squares: the detected three different semi-regular period spacing values of $ g $ modes. Red circle: $ 4f_\mathrm{orb} $ harmonic of orbital frequency. The marker sizes are associated to the mode amplitude as 2$\times 10^{3}\sqrt{A}$.}
\end{figure}
\MakeTable{lcccccc}{8.5cm}{The detected $ g $ modes with period spacings}
{\hline	
	$f_\mathrm{i}$  & $f$      & $A$       & $\Pi$ & $\Delta \Pi$    & $\Delta \Pi$ & SNR \\
	                & d$^{-1}$ & $\mu$mag  & d     & s               &   d          &  \\
	\hline
	$f_\mathrm{4}$  &  2.841  &  1193.73 &  0.352  &                 &       & 14.07\\
	$f_\mathrm{8}$  &  2.723  &  546.94  &  0.367  &  1318.7$\pm$5.2 & 0.015 & 9.29\\[5px]
	$f_\mathrm{15}$ &  2.413  &  207.72  &  0.414  &                 &       & 5.03\\
	$f_\mathrm{5}$  &  2.306  &  1018.45 &  0.434  &  1665.3$\pm$9.5 & 0.019 & 16.39\\
	$f_\mathrm{2}$  &  2.211  &  1412.35 &  0.452  &  1600.6$\pm$4.5 & 0.018 & 14.82\\[5px]
	$f_\mathrm{3}$  &  2.046  &  1317.24 &  0.489  &                 &       & 18.42\\
	$f_\mathrm{1}$  &  1.976  &  1587.18 &  0.506  &  1486.9$\pm$5.3 & 0.017 & 14.10\\
	\hline\multicolumn{7}{p{8.5cm}}{Note: $\Pi$ is the period of the mode. The mean period spacing is $\Delta \Pi_\mathrm{mean}=1517.92\pm131.54$ s (0.017 d).}}
Table~A2 shows all the significant frequencies detected in this region. Most of the $g$ modes appear in the interval of 0.6-3.03 d$^{-1}$. In order to make the list of possibly real gravity modes as reliable as possible, we identified any frequency that is either a linear combination of two more dominant, independent frequencies (chosen as parent frequencies) or a harmonic of one of them. To identify the parent frequencies (Kurtz et al. 2015), we focused on the ones with both SNR and amplitudes above the mean values (SNR$_\mathrm{mean}$ = 9.37 and $A_\mathrm{mean}$ = 621 $\mu$mag). We adopted $f_\mathrm{1}$ to $f_\mathrm{5}$ as parent frequencies (Table A2). The known orbital frequency ($f_\mathrm{orb} = 0.64143\pm0.00015$ d$^{-1}$) was also considered as a parent frequency. We detected a possible coupling between two orbital harmonics ($f_\mathrm{13} = 3f_\mathrm{orb}$ and $f_\mathrm{9} = 4f_\mathrm{orb}$) and the $g$ modes. $f_\mathrm{12}$, $f_\mathrm{16}$, $f_\mathrm{17}$ and $f_\mathrm{18}$ are combinations of these orbital harmonics with one of the parent frequencies, in the form $nf_\mathrm{i}\pm mf_\mathrm{j}$ where $n,~m = [\frac{1}{3},~\frac{1}{2},~1,~2,~3]$ (see Table~A2).\\
Our analysis resulted in seven independent $g$ modes showing a semi-regular spacing with period values ranging from 1318 s (minimum) to 1600 s (maximum) (Table~1). Consequently, the mean of the four period spacings is $\Delta \Pi_\mathrm{mean} = 1517.92\pm131.54$ s. The quoted error is the standard deviation of the different spacings (they are consistent with each other within the adopted uncertainties). Fig. 3 shows the detected frequencies (top panel) and periods (middle panel) along with the semi-regular period spacing pattern (bottom panel left).
In addition, we found that the frequency difference between $f_\mathrm{3}$ and $f_\mathrm{2}$ and between  $f_\mathrm{20}$ and $f_\mathrm{16}$ ($\Delta f = 0.1617\pm0.0009$ d$^{-1}$) is close to $\frac{1}{4}f_\mathrm{orb}$. We also recall that $f_\mathrm{20}$ and $f_\mathrm{16}$ are not independent (Table~A2). Thus, the repeated frequency spacing could be fortuitous.\\
The period-\'{e}chelle diagram for $g$ modes is presented in Fig. 3 (right panel). The periods of all significant frequencies in this region were plotted versus period modulo the mean period spacing (grey circles in the right panel of Fig. 3). Using a step of 1 s around $\Delta \Pi_\mathrm{mean}$, we found the "best" period spacing value of $\Delta \Pi = 1521$ s. The period-\'{e}chelle diagram shows that the detected $ g $ modes (black squares) are settled along a ridge with wiggles. The marker size is associated to the mode amplitude (2$\times 10^{3}\sqrt{A}$). The red circle refers to $4f_\mathrm{orb}$ in the $g$ mode region. The calculated pulsation constant for the $ g $ modes ranges from 0.22 to 0.37 d (Table A2).\\
Lee (2016) found six $g$ modes that are identical to ours, plus some higher orbital harmonics in combination with them. Note that he stopped the analysis using the limit criterion of SNR$\sim$4 (at $f_{30}$ in his Table~3), whereas we find SNR$\sim$8 for the same frequency (at $f_{13} = 3f_{orb}$ in Table~A2). 
\subsection{The high-frequency region}
\begin{figure}[htb]
\includegraphics[width=\columnwidth]{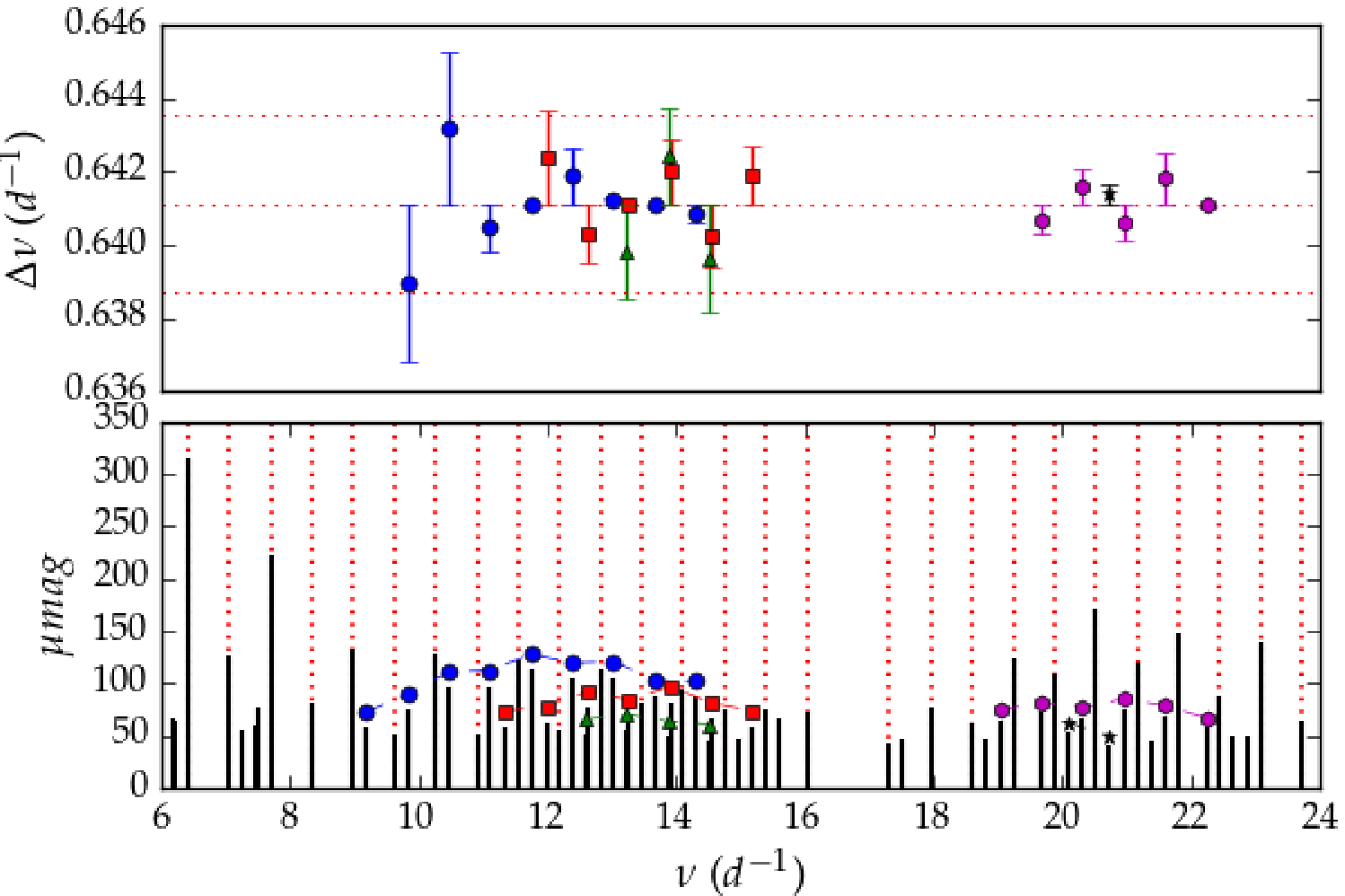}
	\FigCap{Detected $p$ modes with the regular mean spacing very close to the orbital frequency ($\Delta \nu_\mathrm{mean} = 0.64113\pm0.00027$ d$^{-1}$ $\simeq f_\mathrm{orb}$). Red dotted lines ("...") in top panel indicate the boundaries of $\Delta \nu_\mathrm{mean}$ and $\Delta \nu_\mathrm{mean}\pm 3\sigma$. Red dotted lines in the bottom panel show the orbital harmonics. For information on color codes and symbols, see Table 2. }
\end{figure}
The frequencies in the high-frequency region are mostly concentrated in two clusters. The dominant frequencies (excluding harmonics of the orbital frequency) for each cluster are $\nu_\mathrm{max_{1}} = 11.7454\pm0.0012$ d$^{-1}$\footnote{The notation for the frequency changed from $f$ to $\nu$, to differentiate between frequencies in $g$ mode and $p$ mode regions. As the frequencies in each region are sorted by their amplitudes, we avoid confusion by using different notations. This notation does not carry any physical meaning.} and $\nu_\mathrm{max_{2}}=20.9600\pm0.0017$ d$^{-1}$. All the information about the significant frequencies in the $p$ mode region \textit{e.g.} their amplitude, phase (and their errors) and SNR are listed in Tables~A4 and A5. We searched for regular patterns in the frequency domain. Fig. 4 shows five regular frequency patterns with a mean spacing of $\Delta \nu_\mathrm{mean} = 0.64113\pm0.00027$ d$^{-1}$. Within the uncertainty, the detected spacings equal the orbital frequency, $f_\mathrm{orb} = 0.64143\pm0.00015$ d$^{-1}$. This fact, namely that the $p$ modes are regularly split by $f_\mathrm{orb}$, hampers the determination of the large frequency spacing $\Delta \nu$. \\ 
Table 2 lists the values of the regular frequency spacings and their scatter around the mean value. Fig. 4 illustrates the deviations from the mean frequency spacing 
for all the patterns (top panel). The red dotted lines show the $\pm$3$\sigma$ uncertainty levels and the mean value itself.\\
We illustrate the \'{e}chelle diagram of all frequencies located in the $p$ mode region, plotted versus the frequency modulo the orbital frequency in Fig. 5. We can easily see that all the significant frequencies in the high-frequency region fit one of the three ridges in this diagram. 
\MakeTable{lcccccccccc}{13cm}{The detected $p$ modes with regular frequency spacing.}
{\hline	
        $\nu_\mathrm{i}$   & $\nu$     & $\epsilon_{\nu}$       & $A$       & $\Delta \nu$& $\epsilon_{\Delta \nu}$ & SNR & $\Pi$    & Q        &$\Pi$/$\Pi_{0}$& comment\\
	                       & d$^{-1}$  & d$^{-1}$        & $\mu $mag & d$^{-1}$    & d$^{-1}$         & 1d$^{-1}$     & d       &  d        &\\
	\hline  
	P1 & \color{blue}$\circ$         &   \\
	$\nu_\mathrm{49}$ &  9.182   & 0.002 & 57.66     &          &       &  4.14 \\
	$\nu_\mathrm{32}$ &  9.821   & 0.002 & 74.41     &  0.639   & 0.002 &  5.03 & 0.102 & 0.064$\pm$0.014 & 1.938 & \\
    $\nu_\mathrm{18}$ &  10.464  & 0.001 & 95.81     &  0.643   & 0.002 &  5.02 & 0.096 & 0.060$\pm$0.013 & 1.819 & \\                               
    $\nu_\mathrm{17}$ &  11.104  & 0.001 & 97.31     &  0.640   & 0.001 &  5.02 & 0.090 & 0.057$\pm$0.013 & 1.714 & \\                               
    $\nu_\mathrm{13}$ &  11.745  & 0.001 & 112.83    &  0.641   & 0.001 &  5.02 & 0.085 & 0.054$\pm$0.012 & 1.621 & \\                               
    $\nu_\mathrm{16}$ &  12.387  & 0.001 & 104.48    &  0.642   & 0.001 &  5.01 & 0.081 & 0.051$\pm$0.011 & 1.537 & \\                               
    $\nu_\mathrm{15}$ &  13.028  & 0.001 & 105.58    &  0.641   & 0.001 &  5.02 & 0.077 & 0.048$\pm$0.011 & 1.461 & \\                               
    $\nu_\mathrm{20}$ &  13.670  & 0.001 & 88.20     &  0.641   & 0.001 &  5.03 & 0.073 & 0.046$\pm$0.010 & 1.393 & \\                               
    $\nu_\mathrm{22}$ &  14.310  & 0.001 & 86.85     &  0.641   & 0.001 &  5.03 & 0.070 & 0.044$\pm$0.010 & 1.330 & \\ [5px]    
        P2 &\color{red}$\square$                                                     \\
        $\nu_\mathrm{47}$ &  11.341  & 0.002 & 58.86     &          &       &  4.13 & 0.088 & 0.056$\pm$0.012 & 1.679&  \\             
        $\nu_\mathrm{45}$ &  11.983  & 0.002 & 62.93     &  0.642   & 0.001 &  4.96 & 0.083 & 0.053$\pm$0.012 & 1.589& \\                               
        $\nu_\mathrm{28}$ &  12.623  & 0.002 & 77.03     &  0.640   & 0.001 &  5.03 & 0.079 & 0.050$\pm$0.011 & 1.508& \\                               
        $\nu_\mathrm{35}$ &  13.264  & 0.002 & 69.18     &  0.641   & 0.001 &  5.03 & 0.075 & 0.048$\pm$0.011 & 1.435&   \\                               
        $\nu_\mathrm{24}$ &  13.906  & 0.001 & 81.65     &  0.642   & 0.001 &  5.03 & 0.072 & 0.045$\pm$0.011 & 1.369&  \\                               
        $\nu_\mathrm{39}$ &  14.547  & 0.002 & 66.37     &  0.640   & 0.001 &  5.03 & 0.069 & 0.043$\pm$0.010 & 1.309&  \\                               
        $\nu_\mathrm{48}$ &  15.189  & 0.002 & 58.58     &  0.642   & 0.001 &  4.96 & 0.066 & 0.041$\pm$0.009 & 1.253&   \\[5px]
        P3 & \color{OliveGreen}$\bigtriangleup$\\                           
        $\nu_\mathrm{57}^*$&  12.590 & 0.002 & 50.78     &          &       &  4.14 & 0.079 & 0.050$\pm$0.011 & 1.512&  \\                               
        $\nu_\mathrm{53}$ &  13.230  & 0.002 & 55.40     &  0.640   & 0.001 &  4.14 & 0.076 & 0.047$\pm$0.011 & 1.439& \\                                                          
        $\nu_\mathrm{59}$ &  13.872  & 0.002 & 48.77     &  0.642   & 0.001 &  4.14 & 0.072 & 0.045$\pm$0.010 & 1.372&  \\                               
        $\nu_\mathrm{64}$ &  14.512  & 0.002 & 45.94     &  0.640   & 0.001 &  4.15 & 0.069 & 0.043$\pm$0.009 & 1.312&  \\ [5px]
        P4 & \color{Mulberry}$\hexagon$\\                              
        $\nu_\mathrm{43}$ &  19.037  & 0.002 & 64.31     &          &       &  5.03 & 0.052 & 0.033$\pm$0.007 & 1.000 & $\ell=0$? \\                               
        $\nu_\mathrm{34}$ &  19.678  & 0.002 & 71.96     &  0.641   & 0.001 &  5.03 & 0.051 & 0.032$\pm$0.007 & 0.967 & $\ell=0$?\\                               
        $\nu_\mathrm{38}^*$& 20.319  & 0.002 & 66.57     &  0.642   & 0.001 &  5.03 & 0.049 & 0.031$\pm$0.007 & 0.937 & $\ell=0$?\\                               
        $\nu_\mathrm{30}$ &  20.960  & 0.002 & 75.45     &  0.641   & 0.001 &  5.02 & 0.047 & 0.030$\pm$0.007 & 0.908 & $\ell=0$?\\                               
        $\nu_\mathrm{36}$ &  21.602  & 0.002 & 68.48     &  0.642   & 0.001 &  5.03 & 0.046 & 0.029$\pm$0.006 & 0.881 & $\ell=0$?\\                               
        $\nu_\mathrm{50}$ &  22.243  & 0.002 & 56.92     &  0.641   & 0.001 &  5.21 & 0.045 & 0.028$\pm$0.006 & 0.856 & $\ell=0$?\\[5px]
        P5 & \textbf{*}\\
        $\nu_\mathrm{54}$   &  20.078  & 0.002   & 52.65  &        &        & 4.96 & 0.050 & 0.031$\pm$0.007 & 0.948 & \\
        $\nu_\mathrm{68}^*$ &  20.720  & 0.002   & 40.14  &  0.641 & 0.001  & 4.20 & 0.048 & 0.030$\pm$0.007 & 0.919 &\\
\hline\multicolumn{11}{p{13cm}}{Note: The mean frequency spacing is $\Delta \nu=~0.64113\pm0.00027$ d$^{-1}$. The asterisk "*" refers to the modes that are combinations of parent frequencies. "P", "$\Pi$" and "$\Pi_0$" are associated to pattern, period of the mode and period of the fundamental radial mode (Stellingwerf 1979), respectively. Clearly, it can be seen from the table that the period ratio of $\nu_\mathrm{43}$ in Pattern 4 corresponds to the expected value for the fundamental radial mode (whereas none of the other ratios are indicative of radial overtones).}}

\begin{figure}[htb]
\begin{center}
\includegraphics[width=0.6\columnwidth]{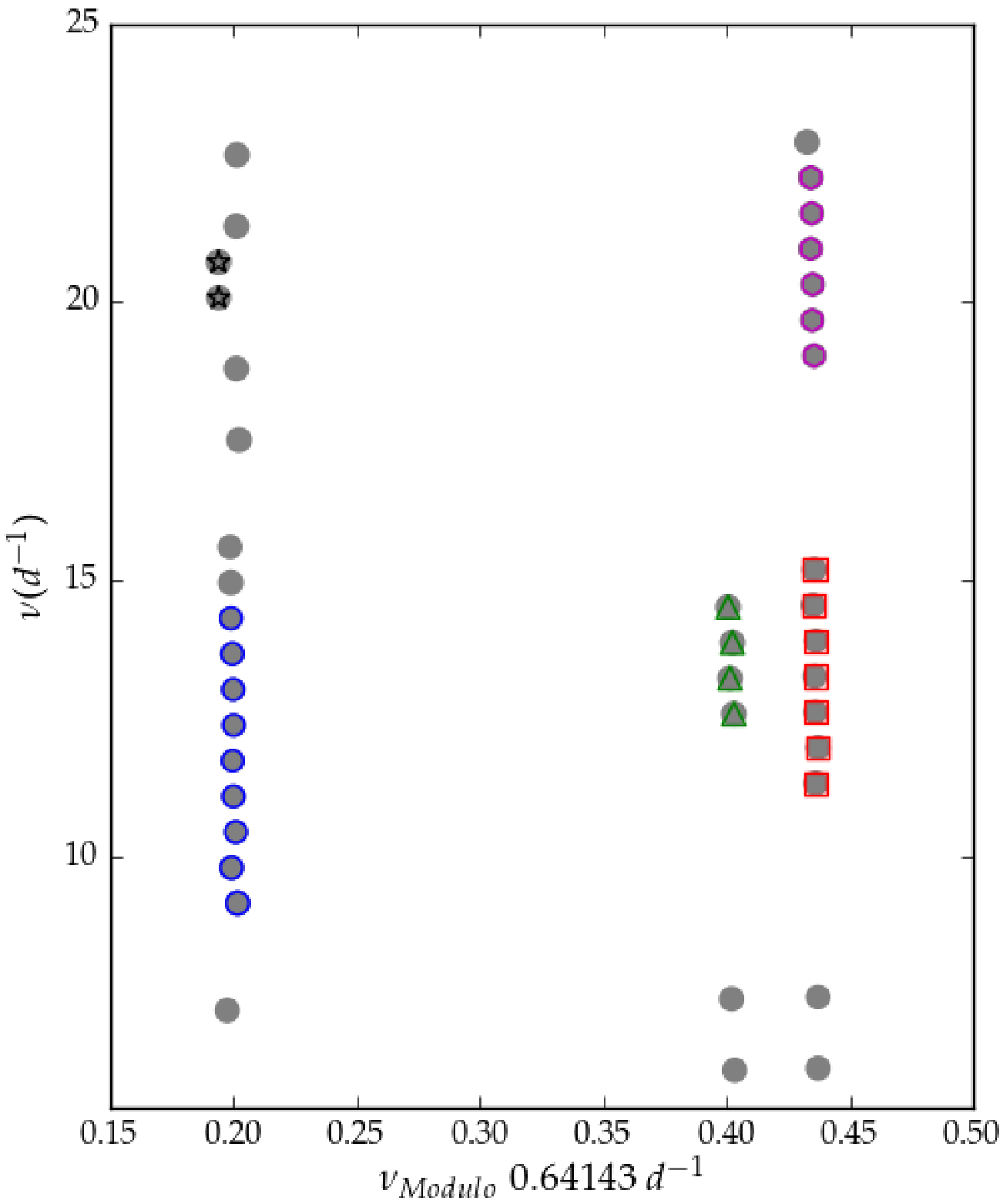}
	\end{center}
	\FigCap{The \'{e}chelle diagram of the frequencies in $p$ mode region that is plotted versus modulo the $ f_\mathrm{orb} = 0.64143$ d$^{-1}$. The grey circles are assigned to all the significant frequencies in the $p$ mode region (Table~A4). Colours follow the colour code of Fig. 4. The marker sizes are associated to the mode amplitude. The detected modes with regular frequency splittings are located in three different ridges. Each ridge is related to different mode degree $\ell$. We can, possibly, associate the right ridge (squares and hexagons) to $\ell=0$ (Table 2 \& Fig. 4), the middle right with triangles to $\ell=2$ and the colored circles in the left ridge to $\ell=1$ modes. Modes with star symbol deviate from locating exactly in this ridge. Furthermore, it is also the left ridge which refers to the frequencies with Q values close to $\delta$ Scuti $p$ modes (Table 2).}
\end{figure}
\subsection{Discussion on the detected high frequencies}
We showed that the shorter pulsation periods (\textit{i.e.} the $p$ modes) are influenced by the orbital frequency, as evidenced by the \'{e}chelle diagram. We also note that this system is circularized. However, we still don't know the true value of the large frequency spacing $\Delta \nu$. It is an interesting fact that similar results were reported by Guo et al.\ (2016) during the pulsation study of KIC~9851944. The latter binary system moves in a circular orbit with an orbital period of 2.61 d. The primary and the secondary components are located inside and near (just outside) the blue edge of the $\gamma$ Dor instability strip, respectively. The authors reported many $p$ modes (in two regions of the periodogram mainly: $4-8$ and $10-15$ d$^{-1}$) which are regularly split by the orbital frequency and explain this effect as being caused by the tidal deformation of the components (\textit{i.e.} ellipsoidal variability). They also reported that different amplitude modulations affect modes of different degree, $\ell$, due to the various cancellation effects. They also detected some low-frequency $g$ modes. Recalling the explanations of Sect.~3.2 and the above mentioned study by Guo et al.\,(2006), we consider that the regular frequency splitting that we detected in the $p$ modes of the high-frequency region is tidal splitting.\\ 
Another possible explanation for the presence of significant frequencies in the $p$ mode region is mode coupling of self-excited $p$ modes (\textit{e.g.} $\nu_\mathrm{max_{1}}=11.7454\pm0.0012$ d$^{-1}$ and $\nu_\mathrm{max_{2}}=20.9600\pm0.0017$ d$^{-1}$) with $g$ modes. However, we found no such coupling 
that could explain the remaining high-frequency modes.\\
We verified whether any of the frequencies were a possible linear combination of parent frequencies. Applying the same approach as in Section~4.2, we chose the most significant frequency of each pattern as a parent frequency (same criteria as in Section~3). We used the following list as parent frequencies: $\nu_\mathrm{27},\nu_\mathrm{13},\nu_\mathrm{24},\nu_\mathrm{30},\nu_\mathrm{54},\nu_\mathrm{53}$. We considered combinations with small integer coefficients \textit{i.e.} 2 to 4. The only combinations found were: $\nu_\mathrm{57} = \nu_\mathrm{54}-\nu_\mathrm{27},~\nu_\mathrm{38}=2\nu_\mathrm{22}-\nu_\mathrm{27},~\nu_\mathrm{68} = \nu_\mathrm{53}+\nu_\mathrm{27} $. These frequencies are located in patterns 3, 4, and 5, respectively, and are indicated with an asterisk "*" in Table~2. \\
\begin{figure}[htb]
\begin{center}
\includegraphics[width=0.7\columnwidth]{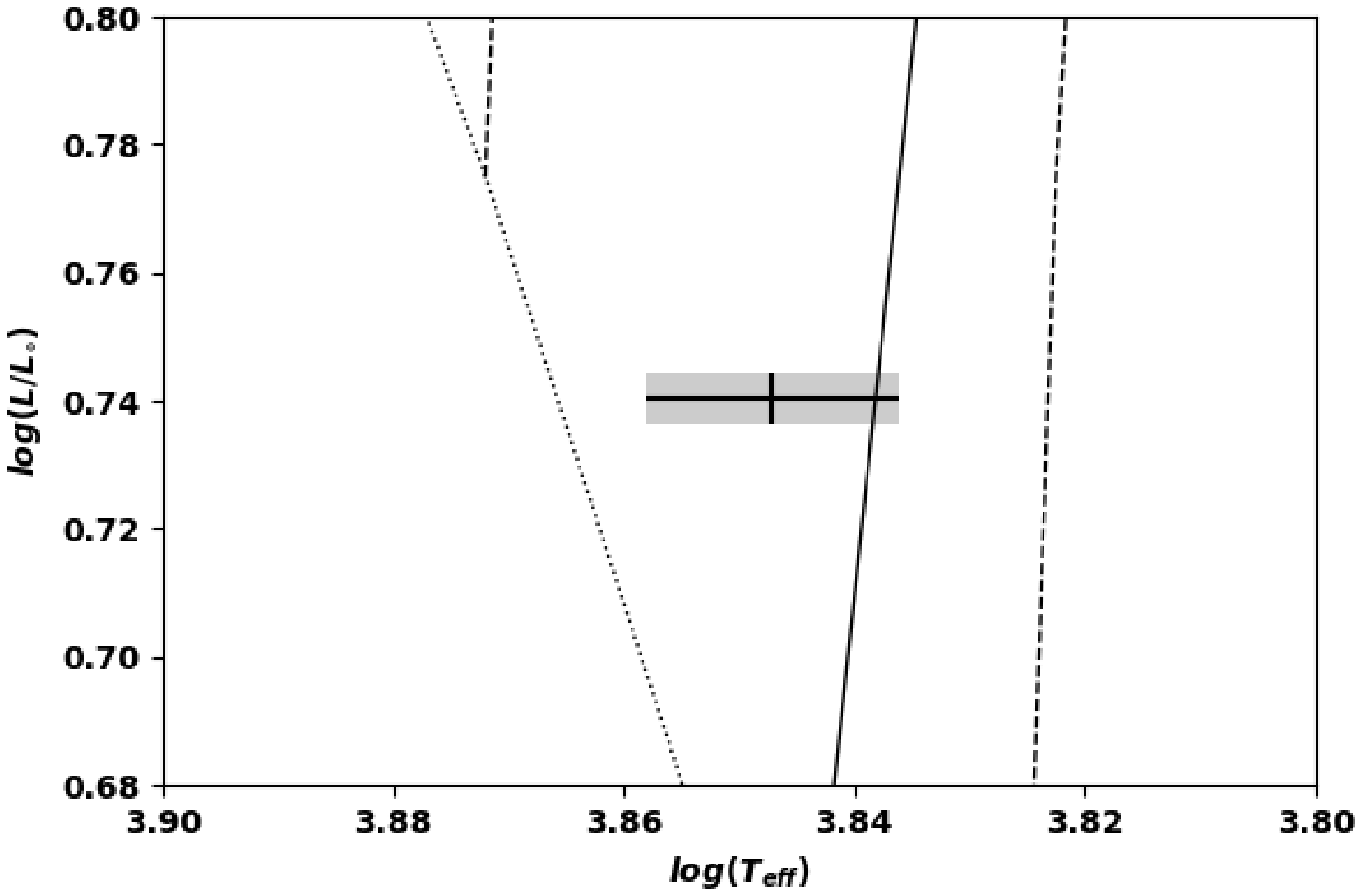}
\end{center}
	\FigCap{A close view to the location of the primary component in the H-R diagram, considering $M_\mathrm{1}=1.55\pm0.11M_{\odot},~R_\mathrm{1}=1.58\pm0.12R_{\odot}$ from Samadi Gh. et al. (2018). The gray rectangular shows the error bars. Dashed lines show theoretical  $\gamma$ Dor IS boundaries (red edge: $\log L/L_{\odot}$ = [0.65,1], $\log {T_{eff}}$ = [3.817,3.825]) for $\ell = 1$ modes and the solid line shows the red edge of the $\delta$ Scuti IS ($\log L/L_{\odot}$ = [0.625,1.3], $\log {T_{eff}}$ = [3.825,3.845]), for the fundamental radial mode and with mixing length parameter $\alpha = 1.8$ (Fig. 15 in Dupret et al. 2005). Pointed line is the ZAMS line. }
\end{figure}

\begin{figure}[htb]
\begin{center}
\includegraphics[width=0.8\columnwidth]{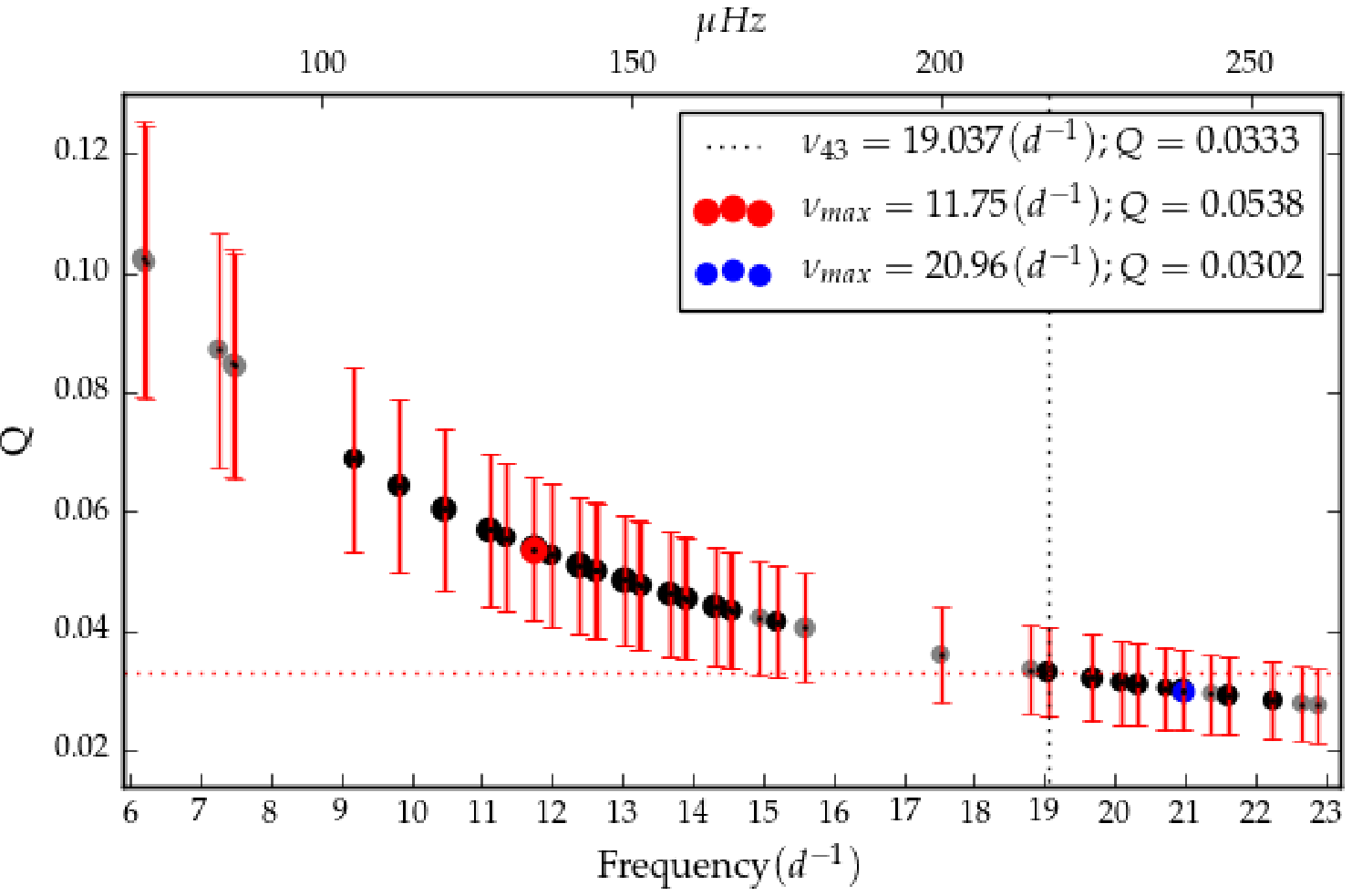}
	\end{center}
	\FigCap{The pulsation constant (Q) for the frequencies in $p$ mode region. Black: Q values (in days) for the frequencies with regular frequency spacing (Table 2). Red and blue circles illustrate the dominant frequencies $\nu_\mathrm{max}$ of two regions in Fourier spectrum where the $p$ modes accumulated. Grey circles: Q values for all the frequencies between 6-24.5 d$^{-1}$ (Tables~A4 and A5, excluding orbital frequencies). The errors plotted here are logarithmic propagation error based on Equation (4) (Table 2). $\nu_\mathrm{43}$ is the $p$ mode with closest Q value to the fundamental radial mode ($p_{1}$, Table 2 \& Fig. 2 Dupret et al. 2005). Marker sizes are associated to the mode amplitude as 5$A\times 10^5$.}
\end{figure}
We repeated the same process with the detrended light curve obtained from the \textit{Kepler Eclipsing Binary Catalog} (KEBC), in order to verify that the chosen detrending method introduces no changes at the level of the detected high-frequency modes. The results show that the high-frequency modes listed in Table~A1 are present and significant in the Fourier Spectrum of the KEBC detrended light curve. This fact proves the presence of the $p$ modes even before subtracting the most adequate binary model.\\ 
Su\'{a}rez et al. (2014) and  Garc{\'{\i}}a Hern{\'a}ndez et al.\ (2015) introduced scaling relations between the (observed) large frequency spacing and the mean density of the star based on a study of $\delta $ Scuti pulsating stars in eclipsing binary systems. These scaling relations are consistent with each other for rotation rates satisfying $\frac{\Omega}{\Omega_\mathrm{K}}\leq0.8$ (with $\Omega_\mathrm{K}$ being the cyclic Keplerian break-up frequency, \footnote{Here $R_{eq}\simeq R_{p}$ (for slow rotators) is adopted in agreement with Garc{\'{\i}}a Hern{\'a}ndez et al. (2015), compared to $R_{eq}\simeq 1.5R_{p}$ (for fast rotators).}$\Omega_\mathrm{K} = \frac{1}{2\pi}\sqrt{\frac{GM}{R_{eq}^{3}}}$. For example, we estimated the large frequency spacing for KIC~6048106 using $\overline{\rho}_\mathrm{prim} = 0.40\pm0.18\overline{\rho}_{\odot},~\Omega_\mathrm{K}=5.44\pm0.8$ d$^{-1}$ (62.67 $\mu$Hz), where $M_\mathrm{1}=1.55\pm0.11M_{\odot},~R_\mathrm{1}=1.58\pm0.12R_{\odot}$ and v$\sin{i}_{1}\simeq$ 44 kms$^{-1}$). Fig.~6 shows that the primary component is located in the overlapping region of the theoretical instability strips for $\delta$ Scuti 
and $\gamma$ Dor stars and near the red edge of $\delta$ Scuti instability strip as modeled by Dupret et al.\ (2005) for $\ell = 1$ modes and with mixing length parameter $\alpha = 1.8$ (Fig. 15 in Dupret et al.\ 2005). Assuming that our system follows the relation by Su\'{a}rez et al.\ (2014), we find $\Delta \nu\simeq$ 5.98 d$^{-1}$ (which is equal to 0.12$\Omega_\mathrm{K}$). This is approximately $10f_\mathrm{orb}$, which is the most dominant frequency in the $p$ mode region (Table A4), and again shows the influence of the tides.\\
Fig.~7 shows the calculated pulsation constant, Q, for all the significant frequencies in the high-frequency region (adopting the physical parameters of the primary component) (Table~A4). This plot illustrates that only the frequencies larger than 19 d$^{-1}$ have Q$\leq$0.033 d. The uncertainties on Q are calculated using the 
law of propagation of the errors based on Equation (4) (Table~2). \\
The pulsation constants, Q, and their uncertainties were calculated according to the derived stellar parameters from the binary modeling (Breger 1990):
\begin{equation}
\label{eq:Q}
Q = \Pi_\mathrm{osc}\sqrt{\frac{\overline{\rho}}{\overline{\rho}_{\odot}}} ,
\end{equation}
where $ \Pi_\mathrm{osc} $ is the oscillation period. As a final step, we compared the period ratios of the detected $p$ modes with those for $\delta$ Scuti-type models (Stellingwerf~1979). $\nu_\mathrm{43}$ = 19.037$\pm$0.002 d$^{-1}$ is the $p$ mode closest to the fundamental radial mode (Q = 0.033$\pm$ 0.007~d). 
\begin{figure}
\begin{center}
\includegraphics[width=0.7\columnwidth]{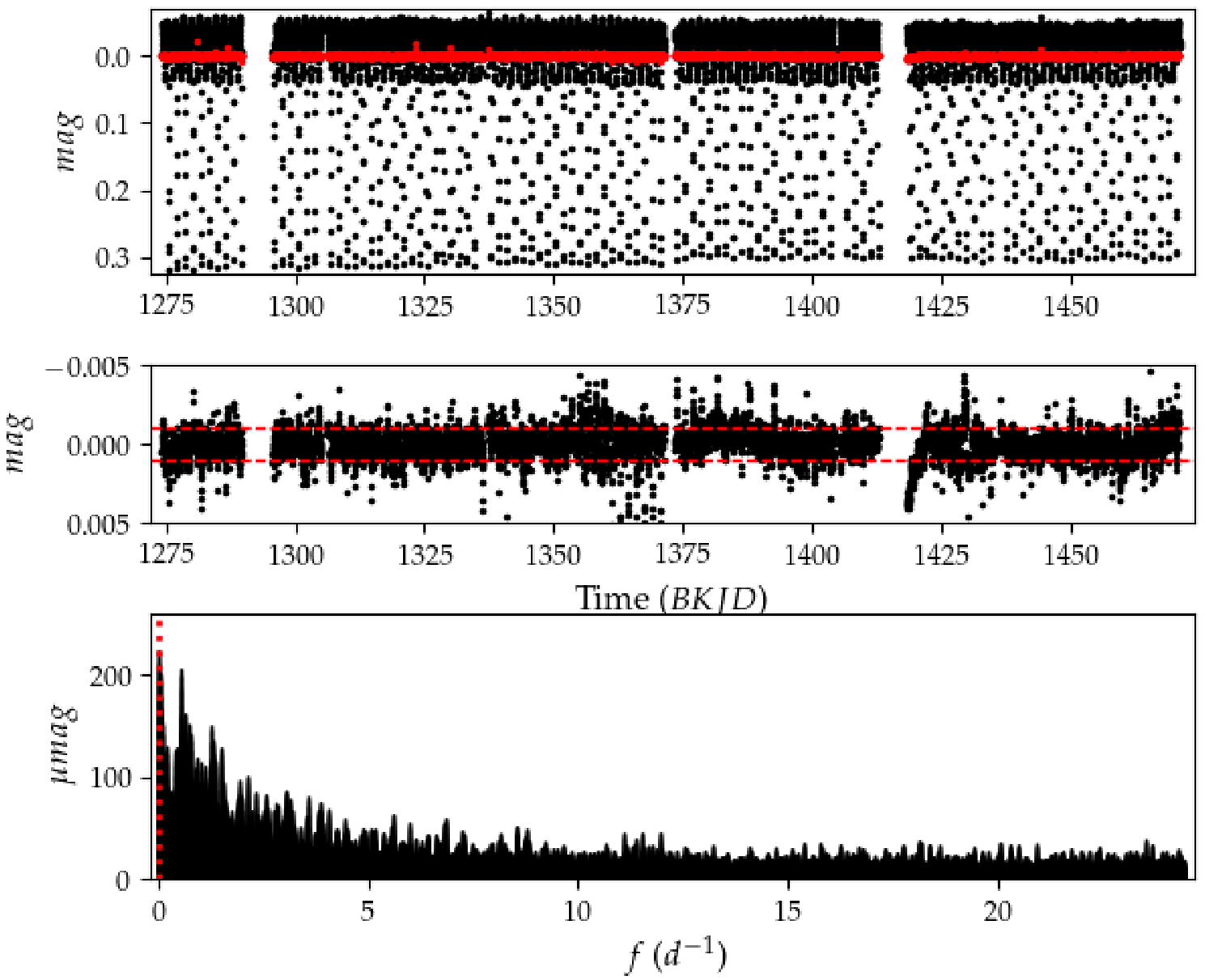}
\end{center}
	\FigCap{Middle panel: The detrended light curve that is prewhitened with the binary and the pulsation signals. Top panel: a comparison of the amplitude of this residual light curve (red) and the detrended observed light curve (black). Bottom panel: The Fourier spectrum of the residual light curve. The frequency highlighted in red (pointed-line) is the only frequency just detectable $ f = 0.0076\pm0.0003$ d$^{-1}$ with SNR $\simeq$4.}
\end{figure}
\subsection{The residuals of the binary and pulsation signals}
After having modeled and subtracted the most significant signals coupled to binarity and pulsation, we checked the final residual light curve for the existence of any signal left. Fig. 8 compares the final residual light curve (middle panel) with the \textit{Kepler} detrended light curve (black in top panel). The red dashed lines show the mean standard deviation of the residuals, $\sigma = 0.0011$ mag (middle panel). This corresponds to a decrease of 85\% in variance (with respect to $\sigma$ = 0.0028 mag, without removal of the pulsation signal). The bottom panel illustrates the Fourier spectrum of the residuals. There is only one frequency that can be considered as marginally significant, that is $f=0.0076\pm0.0003$ (d$^{-1}$), $A=243.14\pm31.06$ ($\mu$mag), $\phi = 0.383\pm0.127~(\deg$), SNR = 3.9. This frequency is seemingly close to the first harmonic of the long-term spot activity of the secondary star ($f_\mathrm{mod} = 0.00359\pm0.00009$ d$^{-1}$, 287$\pm$7 d) derived from the ETV Fourier analysis (Section~2). 
\section{Summary and Conclusions}
We presented the results of a detailed analysis of the \textit{Kepler} light curve of the semi-detached eclipsing binary KIC 6048106 with $ P_\mathrm{orb}$ = 1.559361 $\pm$ 0.000036 d. The fundamental stellar parameters of the components along with an approximate 290-day cycle of spot activity were reported by Samadi Gh. et al.\,(2018). These stellar quantities are the same as the ones derived by Lee (2016). However, a different spot model was assumed. After subtraction of a most plausible \textit{binary with spot model}, we extracted the pulsation frequencies from the residual \textit{Kepler} light curve.  
The significant frequencies appear mostly in three different intervals of the Fourier spectrum: (1) the low-frequency region, $1.96-2.85$ d$^{-1}$, including seven independent $g$ modes with amplitudes between $200-1588~\mu$mag, all of which were reported by Lee (2016); (2) the high-frequency region, mainly 7.49-15.2 d$^{-1}$ and 19-22.5 d$^{-1}$, including 34 modes with amplitudes between $50-106~\mu$mag (but not reported by Lee (2016)). The most dominant frequencies in these two regions are $\nu_\mathrm{max_{1}} = 11.7454\pm0.0012$ d$^{-1}$ and $\nu_\mathrm{max_{2}}=20.9600\pm0.0017$ d$^{-1}$. \\ 
Furthermore, we discovered a semi-regular period spacing pattern for the seven $g$ modes with a mean spacing of $\Delta \Pi_\mathrm{mean} = 1517.92\pm131.54$ s. This may indicate that all seven modes are truly independent frequencies. The period-\'{e}chelle diagram (modulo the mean period spacing) shows that the detected gravity modes are located in a nearly vertical ridge with some wiggles. Four frequencies of this region are combinations of the orbital frequency (mainly 2$ ^\mathrm{nd} $ and 3$ ^\mathrm{rd} $ harmonics). Compared to Lee (2016), our frequency analysis of the residual \textit{Kepler} light curve also revealed various low-amplitude pressure modes.\\
We found that 25 out of 34 detected modes form multiplets that are equidistantly split by the orbital frequency ($\Delta \nu_\mathrm{mean} = 0.64113\pm0.00027$ d$^{-1}$). This feature was already observed in another binary system with a circular orbit (Guo et al.\ 2016). This situation severely hampers the identification of the large frequency spacing in the $p$ mode region. Adopting the derived stellar properties of the primary component ($T_\mathrm{eff}=7033\pm187$ K; $M_\mathrm{bol}=2.90\pm0.01$ mag), its position in the H-R diagram is at the intersection of the $\gamma$ Dor and $\delta$ Scuti instability strips (within the uncertainties). Thus, the primary star KIC~6048106~A is the prime candidate for the detected pulsations. Under that assumption, the Q-values of the regularly spaced $p$ modes in the region 19-22.5 d$^{-1}$ lie between 0.028 and 0.033 d. The Q-value of $\nu_{43} = 19.037\pm$0.002 d$^{-1}$ suggests the presence of the fundamental radial mode.\\
In summary, we found evidence for hybrid pulsations in KIC~6048106~A, though with a clear dominance of the gravity modes. Furthermore, since the equilibrium tide of a close binary system with a circular orbit leads to equidistant frequency splitting of an excited mode by the orbital frequency according to theoretical predictions (Reyniers \& Smeyers 2003a,b, Smeyers 2005), the here described frequencies can be interpreted as tidally split $p$ modes, similar to the conclusion of Guo et al. (2016). In the case of 14~Aur~A, both tidal and rotational splittings as well as the dependence of the apparent amplitude of the pulsation on orbital phase were detected (Webbink et al.\, 2001). We may ask the question why we didn't find any tidal splitting of the more dominant $g$ modes. One reason could be that the effects of tidal distortion are more important in the outer stellar atmosphere where the $p$ modes propagate than in the stellar interior where the $g$ modes are excited.\\ 
We suggest that a follow-up spectroscopic study allowing to refine both the binary and the seismic modeling would be worthwhile for this \textit{Kepler} Algol-type system with its intriguing hybrid pulsation properties. However, the system is faint and multi-epoch spectroscopic observations of high quality are therefore not easily acquired. More systems like this one need to be studied, since the effects of tides on the pulsations of components of close binaries are currently only poorly understood.  
\Acknow{All of the data presented in this paper were obtained from the Mikulski Archive for Space Telescopes (MAST). STScI is operated by the Association of Universities for Research in Astronomy, Inc., under NASA contract NAS5-26555. Support for MAST for non-HST data is provided by the NASA Office of Space Science via grant NNX09AF08G and by other grants and contracts.\\
We gratefully acknowledge a 1-month visitor's grant from the non-profit association "ASBL-VZW Dynamics of the Solar System", and the hospitality of the Royal Observatory of Belgium (ROB), Brussels, Belgium, for the first author. Furthermore, special thanks go to \textit{Dr. Peter De Cat, Dr. Tim Van Hoolst} (ROB) and \textit{Dr. Daniel Reese} (Observatory of Paris, Lesia) for enlightening discussions. We also thank \textit{Paul Van Cauteren} from the Humain Observatory (ROB) for his continuous support. In addition, we like to thank the anonymous referee for many useful comments.}
\begin{appendix}
\section{Full information on significant frequencies in low- and high-frequency regions} \label{sec:full_mode}
This section provides all the information concerning the significant frequencies detected from Fourier analysis of the detrended and residual light curve (Fig. 1). $f\pm \epsilon_{\mathrm f}$ (d$^{-1}$) are the frequencies, $A\pm\epsilon_{\mathrm A} $ their amplitudes ($\mu$ mag) and $\phi\pm\epsilon_{\mathrm \phi}$ phases  ($2\pi$/$\mathrm rad$ ) for the three different regions of the Fourier spectrum: 0.6-3.03, 3.03-6, and 6-24.47 d$^{-1}$. \\
We also provide the information concerning the combination frequencies, harmonics of the orbital frequency, and pulsation constants of the independent modes ($g$ and $p$ modes). 
\setcounter{table}{0}
\renewcommand{\thetable}{A\arabic{table}}
\MakeTable{lcccccc}{12.5cm}{The list of dominant prewhitened frequencies in the Fourier spectrum of detrended Light curve (excluding harmonics of the orbital frequency)}
{\hline	
$f$         & $A $      & $\phi$                 & $\epsilon_{\mathrm f}$ & $\epsilon_{\mathrm A}$ & $\epsilon_{\mathrm \phi}$ & SNR   \\    
 d$^{-1}$   & $\mu $mag & 2$\pi$/$\mathrm rad$   & d$^{-1}$               & $\mu$ mag              & 2$\pi$/$\rm rad$  & 1d$^{-1}$ \\                                      
\hline
 1.97645  &  1600.81  &  0.0285  &  0.00017  &  99.070   &  0.0619  &  14.38  \\
 2.21146  &  1418.77  &  0.0988  &  0.00015  &  77.730   &  0.0548  &  18.30  \\
 2.04601  &  1319.73  &  0.3776  &  0.00020  &  92.080   &  0.0698  &  14.58  \\
 2.84136  &  1192.09  & -0.3045  &  0.00027  &  114.010  &  0.0956  &  14.31  \\
 2.30598  &  1025.98  & -0.0064  &  0.00019  &  69.540   &  0.0678  &  16.13  \\
 2.72332  &  542.97   & -0.1690  &  0.00030  &  57.420   &  0.1058  &  10.14  \\
 22.00644 &  86.27    & -0.1627  &  0.00184  &  56.810   &  0.6585  &  8.02   \\
 20.72406 &  61.09    & -0.4562  &  0.00190  &  41.490   &  0.6793  &  5.23   \\
 23.28883 &  55.91    &  0.1076  &  0.00199  &  39.760   &  0.7111  &  5.26   \\
 22.64694 &  46.14    & -0.0511  &  0.00211  &  34.720   &  0.7526  &  4.25   \\
 16.87559 &  44.61    &  0.0366  &  0.00219  &  34.840   &  0.7810  &  4.25   \\
 21.36400 &  43.67    &  0.4054  &  0.00224  &  34.910   &  0.7995  &  4.26   \\
 18.15945 &  34.59    &  0.2915  &  0.00213  &  26.280   &  0.7599  &  4.25   \\
 \hline\multicolumn{7}{p{9cm}}{Note: The frequencies are sorted according to decreasing amplitude.}}
\setcounter{table}{1}
\renewcommand{\thetable}{A\arabic{table}}
\MakeTable{lcccccccccl}{15cm}{Significant frequencies in the $g$ mode region of the Fourier spectrum of the residuals light curve.}
{\hline
  $f_{\mathrm i}$& $f$        &$A $      &  $\phi$               &$\epsilon_{\mathrm f}$ &$\epsilon_{\mathrm A}$ &  $\epsilon_{\mathrm \phi}$ & SNR   & $\Pi$ & Q    & comment\\    
                 & d$^{-1}$   &$\mu $mag &  2$\pi$/$\mathrm rad$ &d$^{-1}$               &$\mu$ mag              & 2$\pi$/$\rm rad$           & 1d$^{-1}$ &  d & d &\\                 
 \hline 
 $f_\mathrm{1}$  &  1.97643  &  1587.18 &  0.057 &  0.00018 &  101.42 &  0.063 &  14.10 & 0.5060  & 0.320$\pm$0.072 & P; PS  \\              
 $f_\mathrm{2}$  &  2.21152  &  1412.35 &  0.008 &  0.00018 &  88.97  &  0.062 &  14.82 & 0.4522  & 0.286$\pm$0.064 & P; PS  \\              
 $f_\mathrm{3}$  &  2.04602  &  1317.24 &  0.353 &  0.00017 &  80.25  &  0.060 &  18.42 & 0.4888  & 0.309$\pm$0.069 & P; PS  \\             
 $f_\mathrm{4}$  &  2.84138  &  1193.73 & -0.319 &  0.00027 &  114.05 &  0.095 &  14.07 & 0.3519  & 0.223$\pm$0.050 & P; PS  \\             
 $f_\mathrm{5}$  &  2.30600  &  1018.45 & -0.037 &  0.00020 &  72.53  &  0.071 &  16.39 & 0.4337  & 0.274$\pm$0.062 & P; PS  \\             
 $f_\mathrm{6}$  &  0.65077  &  889.49  & -0.006 &  0.00018 &  57.18  &  0.064 &  9.73  &         &                 & $f_\mathrm{orb}+0.0095$   \\                                      
 $f_\mathrm{7}$  &  0.62866  &  671.74  &  0.456 &  0.00020 &  48.24  &  0.071 &  8.15  &         &                 &$f_\mathrm{4}-f_\mathrm{2}$\\                               
 $f_\mathrm{8}$  &  2.72327  &  546.94  & -0.102 &  0.00037 &  72.60  &  0.132 &  9.29  & 0.3672  & 0.232$\pm$0.052 & PS     \\
 $f_\mathrm{9}$  &  2.56515  &  406.52  & -0.067 &  0.00040 &  57.48  &  0.141 &  8.09  &         &                 & $4f_\mathrm{orb}$\\
 $f_\mathrm{10}$ &  3.03248  &  403.69  & -0.346 &  0.00044 &  63.82  &  0.158 &  9.70  &         &                 & $f_\mathrm{3}+1/2f_\mathrm{1}$ \\
 $f_\mathrm{11}$ &  1.27428  &  362.98  & -0.078 &  0.00043 &  55.39  &  0.152 &  8.09  &         &                 & $f_\mathrm{2}-1/3f_\mathrm{4}$ \\
 $f_\mathrm{12}$ &  0.66472  &  336.19  &  0.181 &  0.00034 &  40.75  &  0.121 &  5.53  &         &                 & $1/2f_\mathrm{1}-1/2f_\mathrm{orb}$\\
 $f_\mathrm{13}$ &  1.92296  &  283.04  & -0.067 &  0.00055 &  56.02  &  0.197 &  8.03  &         &                 & $3f_\mathrm{orb}$\\
 $f_\mathrm{14}$ &  1.25665  &  211.18  & -0.359 &  0.00046 &  34.79  &  0.164 &  4.14  &         &                 & $2f_\mathrm{4}-2f_\mathrm{2}$    \\
 $f_\mathrm{15}$ &  2.41326  &  207.71  &  0.408 &  0.00061 &  44.90  &  0.216 &  5.03  & 0.4144  & 0.262$\pm$0.059 & PS    \\
 $f_\mathrm{16}$ &  1.85844  &  193.21  &  0.027 &  0.00066 &  45.27  &  0.234 &  5.03  &         &                 & $2f_\mathrm{2}-4f_\mathrm{orb}$   \\
 $f_\mathrm{17}$ &  2.68708  &  158.45  &  0.185 &  0.00078 &  44.15  &  0.278 &  5.03  &         &                 & $1/3f_\mathrm{5}+3f_\mathrm{orb}$ \\
 $f_\mathrm{18}$ &  2.65393  &  153.27  & -0.281 &  0.00081 &  44.17  &  0.288 &  5.03  &          &                & $3f_\mathrm{orb}+1/3f_\mathrm{2}$ \\
 $f_\mathrm{19}$ &  2.11222  &  151.50  & -0.239 &  0.00069 &  37.12  &  0.245 &  4.13  &          &                & $2f_\mathrm{2}-f_\mathrm{5}$      \\
 $f_\mathrm{20}$ &  1.69674  &  143.64  & -0.113 &  0.00067 &  34.33  &  0.239 &  4.15  &          &                 & $1/3f_\mathrm{4}+1/3f_\mathrm{5}$ \\
\hline\multicolumn{11}{p{15cm}}{Note: P: parent frequency, PS: refers to detected $g$ modes with a regular period spacing pattern. Q is the calculated pulsation constant for the detected $ g $ modes. $f_\mathrm{orb} = 0.64143\pm0.00015$ d$^{-1}$. }}

\setcounter{table}{2}
\renewcommand{\thetable}{A\arabic{table}}
\MakeTable{lccccccl}{10cm}{Significant frequencies of the Fourier spectrum of the residuals light curve, between 3-6 d$^{-1}$.}
{\hline
  $f$       &$A $      &  $\phi$               &$\epsilon_\mathrm{f}$ &$\epsilon_\mathrm{ A}$ &  $\epsilon_\mathrm {\phi}$ & SNR     & comment\\
  d$^{-1}$  &$\mu$ mag &  2$\pi$/$\mathrm rad$ &d$^{-1}$              &$\mu mag$              & 2 $\pi$/$\mathrm rad$      & 1d$^{-1}$ &      \\
  \hline
  3.84774  &  425.55  &  0.073  &  0.00073  &  111.21  &  0.261  &  14.09  & $6f_\mathrm{orb}$ \\
  5.13045  &  395.53  & -0.135  &  0.00080  &  113.56  &  0.287  &  14.08  & $8f_\mathrm{orb}$ \\
  4.48886  &  243.57  &  0.179  &  0.00083  &  72.19   &  0.296  &  9.29   & $7f_\mathrm{orb}$ \\
  3.20532  &  210.70  & -0.491  &  0.00063  &  47.10   &  0.223  &  5.02   & $5f_\mathrm{orb}$ \\
  5.77160  &  184.96  & -0.019  &  0.00097  &  64.01   &  0.346  &  9.73   & $9f_\mathrm{orb}$ \\ 
 \hline\multicolumn{8}{p{10cm}}{}}                                                                                                           

\setcounter{table}{3}
\renewcommand{\thetable}{A\arabic{table}}
\MakeTable{lcccccccl}{11cm}{Significant frequencies in $p$ modes region of the Fourier spectrum of the residuals light curve.}{\hline
 $\nu_\mathrm{i}$ & $\nu$    & $A$      & $\phi$     & $\epsilon_{\nu}$   & $\epsilon_{\rm A}$ & $\epsilon_{\phi}$ & SNR   & Comment \\
                  & d$^{-1}$ & $\mu$mag & 2$\pi$/rad & d$^{-1}$           & $\mu$mag           & 2$\pi$/rad        & 1d$^{-1}$ &     \\
 \hline         
 $\nu_\mathrm{1} $ &  6.41322   &  314.58  & -0.462  &  0.00100  &  112.79 &  0.358 & 14.09&  $10f_\mathrm{orb}$ \\ 
 $\nu_\mathrm{2} $ &  7.69581   &  221.91  &  0.482  &  0.00094  &  74.34  &  0.335 & 9.29 &  $12f_\mathrm{orb}$ \\ 
 $\nu_\mathrm{3} $ &  20.52136  &  172.21  &  0.229  &  0.00121  &  74.26  &  0.431 & 9.29 &  $32f_\mathrm{orb}$ \\ 
 $\nu_\mathrm{4} $ &  21.80426  &  148.76  & -0.190  &  0.00137  &  72.86  &  0.489 & 9.29 &  $34f_\mathrm{orb}$ \\ 
 $\nu_\mathrm{5} $ &  23.08655  &  140.16  &  0.198  &  0.00146  &  72.89  &  0.520 & 9.29 &  $36f_\mathrm{orb}$ \\ 
 $\nu_\mathrm{6} $ &  8.97830   &  133.74  & -0.492  &  0.00120  &  57.32  &  0.428 & 8.09 &  $14f_\mathrm{orb}$ \\   
 $\nu_\mathrm{7} $ &  10.26027  &  128.80  &  0.283  &  0.00125  &  57.47  &  0.446 & 8.09 &  $16f_\mathrm{orb}$ \\
 $\nu_\mathrm{8} $ &  7.05393   &  126.34  &  0.395  &  0.00105  &  47.55  &  0.376 & 5.02 &  $11f_\mathrm{orb}$ \\
 $\nu_\mathrm{9} $ &  19.23855  &  123.34  & -0.477  &  0.00130  &  57.08  &  0.462 & 8.08 &  $30f_\mathrm{orb}$ \\
 $\nu_\mathrm{10}$ &  11.54343  &  121.62  &  0.451  &  0.00111  &  48.28  &  0.396 & 5.02 &  $18f_\mathrm{orb}$ \\
 $\nu_\mathrm{11}$ &  21.16253  &  120.43  &  0.401  &  0.00134  &  57.52  &  0.477 & 8.08 &  $33f_\mathrm{orb}$ \\
 $\nu_\mathrm{12}$ &  12.82602  &  114.22  &  0.418  &  0.00115  &  46.74  &  0.409 & 5.01 &  $20f_\mathrm{orb}$ \\
 $\nu_\mathrm{13}$ &  11.74543  &  112.83  & -0.214  &  0.00121  &  48.78  &  0.432 & 5.02 &  P; FS  \\
 $\nu_\mathrm{14}$ &  19.88027  &  110.01  & -0.045  &  0.00146  &  57.30  &  0.520 & 8.09 &  $31f_\mathrm{orb}$ \\
 $\nu_\mathrm{15}$ &  13.02851  &  105.58  &  0.081  &  0.00126  &  47.34  &  0.448 & 5.02 &  \\
 $\nu_\mathrm{16}$ &  12.38730  &  104.48  & -0.023  &  0.00127  &  47.45  &  0.454 & 5.01 &  \\
 $\nu_\mathrm{17}$ &  11.10434  &  97.31   & -0.476  &  0.00133  &  46.10  &  0.473 & 5.02 &  FS   \\
 $\nu_\mathrm{18}$ &  10.46385  &  95.81   &  0.381  &  0.00140  &  47.85  &  0.499 & 5.02 &  FS    \\
 $\nu_\mathrm{19}$ &  14.10777  &  94.42   & -0.359  &  0.00139  &  46.86  &  0.496 & 5.02 &  $22f_\mathrm{orb}$ \\
 $\nu_\mathrm{20}$ &  13.66960  &  88.20   &  0.377  &  0.00143  &  44.93  &  0.509 & 5.03 &  \\
 $\nu_\mathrm{21}$ &  22.44521  &  87.43   &  0.286  &  0.00150  &  46.93  &  0.536 & 5.01 &  $35f_\mathrm{orb}$ \\
 $\nu_\mathrm{22}$ &  14.31046  &  86.85   & -0.049  &  0.00145  &  45.09  &  0.519 & 5.03 &  \\
 $\nu_\mathrm{23}$ &  13.46932  &  82.07   & -0.418  &  0.00151  &  44.20  &  0.538 & 5.03 & $21f_\mathrm{orb}$\\
 $\nu_\mathrm{24}$ &  13.90641  &  81.65   & -0.064  &  0.00155  &  45.16  &  0.553 & 5.03 & P; FS    \\
 $\nu_\mathrm{25}$ &  8.33578   &  80.56   &  0.469  &  0.00149  &  42.92  &  0.532 & 5.21 & $13f_\mathrm{orb}$ \\
 $\nu_\mathrm{26}$ &  17.95508  &  77.96   & -0.267  &  0.00170  &  47.51  &  0.609 & 5.02 & $28f_\mathrm{orb}$ \\
 $\nu_\mathrm{27}$ &  7.49290   &  77.38   &  0.266  &  0.00148  &  41.12  &  0.531 & 4.96 & P                  \\
 $\nu_\mathrm{28}$ &  12.62330  &  77.03   & -0.315  &  0.00164  &  45.32  &  0.588 & 5.03 & $3\nu_\mathrm{15}-2\nu_\mathrm{53}$; FS\\
 $\nu_\mathrm{29}$ &  15.39029  &  76.01   & -0.191  &  0.00165  &  44.84  &  0.589 & 5.03 & $24f_\mathrm{orb}$  \\
 $\nu_\mathrm{30}$ &  20.96004  &  75.45   &  0.285  &  0.00173  &  46.57  &  0.617 & 5.02 & P; FS   \\
 $\nu_\mathrm{31}$ &  14.75178  &  74.57   & -0.218  &  0.00167  &  44.48  &  0.596 & 5.03 & $23f_\mathrm{orb}$  \\
 $\nu_\mathrm{32}$ &  9.82064   &  74.41   &  0.113  &  0.00169  &  44.79  &  0.601 & 5.03 &                \\
 $\nu_\mathrm{33}$ &  16.03459  &  72.55   &  0.475  &  0.00174  &  45.08  &  0.621 & 5.03 & $24f_\mathrm{orb}$ \\
 $\nu_\mathrm{34}$ &  19.67780  &  71.96   & -0.209  &  0.00178  &  45.73  &  0.635 & 5.03 & FS  \\
 $\nu_\mathrm{35}$ &  13.26441  &  69.18   & -0.091  &  0.00183  &  45.24  &  0.654 & 5.03 & FS  \\
 $\nu_\mathrm{36}$ &  21.60186  &  68.48   & -0.476  &  0.00188  &  46.03  &  0.672 & 5.03 & FS  \\
 $\nu_\mathrm{37}$ &  15.59298  &  67.10   &  0.068  &  0.00187  &  44.96  &  0.670 & 5.03 &     \\
 $\nu_\mathrm{38}$ &  20.31940  &  66.57   &  0.353  &  0.00189  &  45.03  &  0.676 & 5.03 & $2\nu_\mathrm{22}-\nu_\mathrm{27}$  \\
 $\nu_\mathrm{39}$ &  14.5467   &  66.37   &  0.378  &  0.00188  &  44.62  &  0.672 & 5.03 & FS \\
 $\nu_\mathrm{40}$ &  6.17622   &  65.53   &  0.006  &  0.00153  &  35.87  &  0.547 & 4.14 & \\
\hline\multicolumn{9}{p{11cm}}{Note: P: parent frequency, FS: refers to the detected $ p $ modes with regular frequency spacings.}}                                                                

\setcounter{table}{4}
\renewcommand{\thetable}{A\arabic{table}}
\MakeTable{lcccccccl}{11cm}{Continued from Table A4.}
{\hline
$\nu_\mathrm{i}$ & $\nu$    & $A$      & $\phi$     & $\epsilon_{\nu}$   & $\epsilon_{\rm A}$ & $\epsilon_{\phi}$ & SNR       &  Comment \\
                 & d$^{-1}$ & $\mu$mag & 2$\pi$/rad & d$^{-1}$           & $\mu$mag           & 2$\pi$/rad        & 1d$^{-1}$ &      \\                   
 \hline
 $\nu_\mathrm{41}$ &  23.72726  &  64.73   &  0.024  &  0.00193  &  44.68  &  0.690 & 5.03 & $37f_\mathrm{orb}$ \\
 $\nu_\mathrm{42}$ &  6.21006   &  64.33   & -0.370  &  0.00149  &  34.34  &  0.533 & 4.15 & \\
 $\nu_\mathrm{43}$ &  19.03710  &  64.31   & -0.041  &  0.00192  &  44.22  &  0.687 & 5.03 & FS \\
 $\nu_\mathrm{44}$ &  18.59781  &  62.97   & -0.199  &  0.00184  &  41.36  &  0.656 & 4.96 & $29f_\mathrm{orb}$       \\
 $\nu_\mathrm{45}$ &  11.98297  &  62.93   &  0.335  &  0.00184  &  41.42  &  0.658 & 4.96 & FS  \\
 $\nu_\mathrm{46}$ &  7.45785   &  59.46   &  0.292  &  0.00162  &  34.45  &  0.579 & 4.15 &                      \\
 $\nu_\mathrm{47}$ &  11.34056  &  58.87   & -0.086  &  0.00173  &  36.44  &  0.619 & 4.13 &                      \\
 $\nu_\mathrm{48}$ &  15.18859  &  58.58   &  0.490  &  0.00197  &  41.23  &  0.703 & 4.96 & FS  \\
 $\nu_\mathrm{49}$ &  9.18168   &  57.66   & -0.209  &  0.00171  &  35.27  &  0.611 & 4.14 & $2\nu_{18}-\nu_{13}$ \\
 $\nu_\mathrm{50}$ &  22.24294  &  56.92   & -0.203  &  0.00211  &  42.83  &  0.752 & 5.21 & FS  \\
 $\nu_\mathrm{51}$ &  7.25316   &  56.27   &  0.484  &  0.00164  &  32.88  &  0.584 & 3.86 &                       \\
 $\nu_\mathrm{52}$ &  12.18789  &  56.15   & -0.090  &  0.00180  &  36.14  &  0.643 & 4.13 & $19f_\mathrm{orb}$    \\
 $\nu_\mathrm{53}$ &  13.23006  &  55.40   &  0.066  &  0.00190  &  37.68  &  0.680 & 4.14 & P; FS  \\
 $\nu_\mathrm{54}$ &  20.07839  &  52.65   & -0.151  &  0.00218  &  40.93  &  0.777 & 4.96 &                       \\
 $\nu_\mathrm{55}$ &  10.90542  &  52.06   & -0.314  &  0.00186  &  34.61  &  0.664 & 4.15 & $17f_\mathrm{orb}$   \\
 $\nu_\mathrm{56}$ &  9.61897   &  51.40   & -0.276  &  0.00188  &  34.58  &  0.672 & 4.15 & $15f_\mathrm{orb}$   \\
 $\nu_\mathrm{57}$ &  12.59021  &  50.78   &  0.022  &  0.00196  &  35.55  &  0.699 & 4.14 & $\nu_\mathrm{54}-\nu_\mathrm{27}$; FS \\
 $\nu_\mathrm{58}$ &  22.65146  &  49.05   &  0.423  &  0.00207  &  36.30  &  0.740 & 4.13 & \\
 $\nu_\mathrm{59}$ &  13.87248  &  48.77   &  0.415  &  0.00207  &  36.02  &  0.738 & 4.14 & FS  \\
 $\nu_\mathrm{60}$ &  22.88285  &  48.24   & -0.284  &  0.00209  &  36.09  &  0.748 & 4.14 & \\
 $\nu_\mathrm{61}$ &  17.52088  &  47.83   & -0.138  &  0.00211  &  36.08  &  0.754 & 4.14 & \\
 $\nu_\mathrm{62}$ &  18.80266  &  47.53   & -0.038  &  0.00204  &  34.66  &  0.729 & 4.15 & \\
 $\nu_\mathrm{63}$ &  14.95183  &  47.30   & -0.176  &  0.00205  &  34.56  &  0.730 & 4.15 & \\
 $\nu_\mathrm{64}$ &  14.51214  &  45.94   & -0.261  &  0.00210  &  34.50  &  0.750 & 4.15 & FS\\
 $\nu_\mathrm{65}$ &  21.36840  &  44.81   &  0.116  &  0.00225  &  35.94  &  0.802 & 4.14 & \\
 $\nu_\mathrm{66}$ &  24.37041  &  42.95   &  0.407  &  0.00232  &  35.53  &  0.827 & 4.14 & $38f_\mathrm{orb}$   \\
 $\nu_\mathrm{67}$ &  17.318157 &  42.10   & -0.017  &  0.00223  &  33.49  &  0.795 & 4.20 & $27f_\mathrm{orb}$  \\
 $\nu_\mathrm{68}$ &  20.71979  &  40.14   & -0.238  &  0.00234  &  33.55  &  0.835 & 4.20 & $\nu_\mathrm{53}+\nu_\mathrm{27}$\\ 
\hline\multicolumn{9}{p{11cm}}{ }}
 
\end{appendix}

\end{document}